\documentclass[a4paper,fleqn,usenatbib]{mnras}

\usepackage[T1]{fontenc}
\usepackage{ae,aecompl}

\usepackage{graphicx}	% Including figure files
\usepackage{amsmath}	% Advanced maths commands
\usepackage{amssymb}

\usepackage{graphicx}	% Including figure files
\usepackage{amsmath}	% Advanced maths commands
\usepackage{amssymb}	% Extra maths symbols
\usepackage{multicol}        % Multi-column entries in tables
\usepackage{bm}		% Bold maths symbols, including upright Greek
\usepackage{pdflscape}	% Landscape pages
\usepackage{hyperref}

\usepackage{graphicx,times}
\usepackage{amsmath}
\usepackage[T1]{fontenc}
\usepackage{aecompl}
\usepackage{caption}
\newcommand{\x}[1]{\text{#1}}
\usepackage{booktabs,caption,fixltx2e}
\usepackage[flushleft]{threeparttable}
\usepackage{tabularx}
\usepackage{pdflscape}
\usepackage{rotating}

\usepackage[export]{adjustbox}
\usepackage{graphics}
\usepackage{amssymb,amsmath}
\usepackage{url}
\usepackage{textcomp}
\usepackage{color}
\usepackage{subcaption}
\usepackage{longtable}
\usepackage{graphicx}

\usepackage{lscape}
\usepackage{float}
\usepackage{rotating}
\usepackage{perpage}

\title[]{Ring Galaxies in the EAGLE Hydrodynamical Simulations}

\author[A.~Elagali et al.]
{\parbox{\textwidth}{Ahmed~Elagali$^{1,2}$\thanks{E-mail  ahmed.elagali@icrar.org},
\ Claudia D. P. Lagos$^{1,2}$,  O. Ivy Wong$^{1,2}$, Lister~Staveley-Smith$^{1,2}$, James W. Trayford$^{3,4}$, Matthieu Schaller$^{3,4}$, Tiantian Yuan$^{2,5}$, Mario G. Abadi$^{6,7}$} \vspace{0.4cm}\\
\parbox{\textwidth}{%$^{1}$Astrophysics, Cosmology and Gravity Centre (ACGC), Astronomy Department, 
$^{1}$International Centre for Radio Astronomy Research (ICRAR), M468, The University of Western Australia, 35 Stirling Highway, Crawley, WA 6009, Australia\\
$^{2}$ARC Centre of Excellence for All Sky Astrophysics in 3 Dimensions (ASTRO 3D)\\
$^{3}$Leiden Observatory, Leiden University, PO Box 9513, NL-2300 RA Leiden, the Netherlands\\
$^{4}$Department of Physics, Institute for Computational Cosmology, University of Durham, South Road, Durham DH1 3LE, UK\\
$^{5}$Centre for Astrophysics and Supercomputing, Swinburne University of Technology, Hawthorn, Victoria 3122, Australia\\
$^{6}$Instituto de Astronom$\acute{i}$a Te$\acute{o}$rica y Experimental, CONICET-UNC, Laprida 854, X5000BGR,C$\acute{o}$rdoba, Argentina\\
$^{7}$Observatorio Astron$\acute{o}$mico de C$\acute{o}$rdoba, Universidad Nacional de C$\acute{o}$rdoba, Laprida 854, X5000BGR, C$\acute{o}$rdoba, Argentina\\}}
\begin{document}

\date{Accepted 00. Received 00; in original form 00}

\pagerange{\pageref{firstpage}--\pageref{lastpage}} \pubyear{2017}

\maketitle

\label{firstpage}

\begin{abstract}

We study the formation and evolution of ring galaxies in the Evolution and Assembly of GaLaxies and their Environments
(EAGLE) simulations. We use the largest reference model Ref-L100N1504, a cubic cosmological volume of $100$ comoving
megaparsecs on a side, to identify and characterise these systems through cosmic time. The number density of ring galaxies 
in EAGLE  is in broad agreement with the observations. The vast majority of ring galaxies identified in EAGLE ($83\,$per cent) 
have an interaction origin, i.e., form when one or more companion galaxies drop-through a disk galaxy. The remainder 
($17\,$per cent) have very long-lived ring morphologies ($> 2\,$Gyr) and host strong bars. Ring galaxies are HI
rich galaxies, yet display inefficient star formation activity and  tend to reside in the green valley  particularly at $z\,\gtrsim\,0.5$. 
This inefficiency is  mainly due to the low pressure and metallicity  of their interstellar
medium (ISM)  compared  with the ISM of similar star-forming galaxies. 
We find that the interaction(s) is responsible for decreasing the ISM pressure by causing the ISM gas to flow from the inner regions to the outer
disk, where the ring feature forms. At a fixed radius, the star formation efficiency of ring galaxies is indistinguishable from their star-forming counterparts, and 
thus the main reason for their integrated lower efficiency is the different gas surface density profiles.
Since galaxy morphologies are not used to tune the parameters in hydrodynamical simulations, the experiment performed here demonstrates 
the success of the  current numerical models in EAGLE.

\end{abstract}


\section{Introduction} 
Galaxy morphology is strongly correlated with the physical properties of galaxies such as their star
formation histories and dynamical structures, and hence provides crucial insights into galaxy formation
and is a key diagnostic of their evolution. In the local Universe, galaxies are categorised into three main morphological types, 
i.e., spirals, spheroids, and irregular dwarf systems \citep[][the Hubble sequence]{1926Hubble}.
Observations of galaxies reveal that a considerable fraction ($\sim10$ per cent) of galaxies in the local 
Universe have irregular morphological structures and do not fit within the conventional Hubble classification \citep{2010Nair,2011Baillard,2013Willett}. 
These morphologically disturbed systems are usually
undergoing interactions and/or mergers with neighbouring galaxies \citep{1977Toomre, 1978White, 1996Barnes, 2006Naab, 2014Naab, 2017Rodriguez-Gomez}.
One of the most peculiar systems in our local Universe are the so-called ring 
galaxies. Ring galaxies are generally divided into two sub-classes, the P and O-types \citep{Few1986}. P-type rings
result from an off-centre passage of a compact companion galaxy, the intruder, through the disk of much
more massive spiral galaxy, the target, see for example \citet{Lynds1976,Theys1977,Struck-Marcell-Lotan1990}. 
This type of collision produces a density wave that radially transports the gas and stars into a ring morphology
throughout the disk of the target galaxy \citep{Appleton1996, Mayya-2005, Wong2006, Romano2008,Fogarty-2011,Parker-2015,conn-2016,Elagali2018}.
%The density of the ring and the perturbation in the target's velocity field depend on the mass of the intruder, i.e,
%high mass intruders produce sharper rings, faster expansions and multiple rings in some cases \citep{ Gerber1996}. 
Depending on the collision's geometry, the target's nucleus can be displaced off its dynamical centre,
the whole system (ring and nucleus) can be slightly dislocated from the previous plane of the disk \citep{Mapelli2012} and 
in some extreme cases the nucleus can completely be disrupted by the collision, producing a centrally smoothed ring with no apparent
nucleus \citep{Madore2009}. The O-type are the resonance ring galaxies \citep{deVaucouleurs-1959, 2015Herrera, Buta-2017}.
These rings are preferentially found in barred galaxies, with the ring encircling the bar and forming the familiar $\theta$-shape.
The O-type rings are not the products of violent galaxy collisions, but rather are formed by gas accumulation at Lindblad resonances
under the continuous influence of gravity torques from the bars.\\

Collisional ring galaxies (P-type) have been widely studied in idealised (isolated) interaction simulations,
in which a companion galaxy collides with a disk galaxy using certain interaction parameters to produce
the ring morphology \citep{Lynds1976,Theys1977,Struck-Marcell-Lotan1990,Hernquist-1993,Gerber1996,1997Athanassoula,2001Horellou, MapelliMoore-2008,2012Smith,Mapelli2012,2018Renaud}.
These simulations show that certain impact  parameters and collision angles can induce warps 
in the disk, affect the morphology and the star formation history of the target disk galaxy \citep{Fiacconi-2012}.
\citet{Mihos-1994} and \citet{Mapelli-2008} showed in their simulations that it takes around $100\,$Myr after the collision to develop the ring in the disk of these galaxies 
and that the ring morphology remains visible for $\sim0.5\,$Gyr, which is very short compared to the Hubble timescale. Since the formation mechanisms of collisional ring galaxies are very well 
constrained via simulations, they can be considered as galaxy-scale perturbation experiments
that facilitate  the study of extreme modes of interaction-triggered star formation and feedback processes
\citep{Higdon-2012, 2015Higdon, 2017Wong,2018Renaud}.\\

Even though non-cosmological isolated interaction simulations are cornerstones in understanding the formation mechanisms of collisional 
ring galaxies, they lack the statistical basis that would allow for a systematic study of these galaxies in the local Universe. 
This is because these simulations consist of only the target and the  intruder companion, using  arbitrary initial 
conditions and collision parameters to induce the ring morphology in the target galaxy. Hence, it is important to expand
the sample of simulated collisional ring galaxies and study larger volumes that cosmologically represent the local Universe.
This will advance our understanding of the collisional ring galaxies' number density, their evolution and whether the drop-through interaction
proposed in non-cosmological isolated interaction simulations is frequent enough to explain the
observed number density of ring galaxies \citep{Lavery-2004, Elmegreen-2006}. Further, observations of collisional ring galaxies suggests that these systems  contain on average high amounts of HI gas in comparison
with galaxies that have the same stellar mass \citep{Elagali2018}, yet are H$_{2}$ deficient \citep{2015Higdon, 2017Wong} especially at
the outer rings where the atomic hydrogen surface density is the highest \citep[e.g., in the Cartwheel galaxy $\Sigma_{HI}\,$=$\,19-65\,$M$_{\odot}\,$pc$^{-2}$;][]{2015Higdon}.
The reason behind this deficiency is not yet well understood, however \citet{2017Wong} and \citet{2015Higdon} hypothesise that the ISM in ring galaxies behaves differently
as a result of the extreme conditions in, e.g., pressure, temperature and metallicity, induced by the drop-through collision. \\

In their N-body cosmological simulation, \citet{DOnghia-2008} adopted a method to identify progenitor merging haloes that
host collisional ring galaxies based on their halo masses, the mass ratio between the two merging haloes, and the 
the impact parameter of the collision (the initial distance vector between the centre of mass of the two merging haloes).
They used the resultant number density of collisional ring galaxies throughout cosmic time to place constraints on the merger rate.
But it is unclear how biased rings are as tracers of interactions and whether that bias is tim1990Strucke invariant or not.
A more recent study of ring galaxies in a cosmological hydrodynamical dynamical simulation was conducted by the Illustris team \citep{Snyder-2015}.
They explored the morphology of galaxies in these simulations and found that at $z\,$=$\,0$ a considerable fraction
of galaxies within the mass range $10^{10.5}-10^{11}\,$M$_{\odot}$ have a distinct ring/C-shape morphology. However, the 
abundance of these systems in their simulations is much higher than observational studies suggest.
One possible reason for this over population of ring galaxies in Illustris is the choice of interstellar medium (ISM) and feedback models.
While this is still inconclusive, in the new Illustris project \citep[Illustris-TNG;][]{2017PillepichNelson,2017Springel,2017Naiman,2017Marinacci,2017NelsonPillepich}
ring-like structures are not as abundant, which suggests that the problem with Illustris is likely a combination of effects from different subgrid physics modules.
This is because in Illustris-TNG the stellar and AGN feedback models are implemented differently in comparison to Illustris, 
which possibly affects the way gas piles up in galaxies and consequently their morphology \citep{2017-Weinberger,2017PillepichNelson}.\\

Here, we intend to study collisional ring galaxies in the EAGLE simulations, trace the abundance of these systems with redshift,
compare it with observations, and study their formation mechanisms. The EAGLE simulations, which stands for Evolution
and Assembly of GaLaxies and their Environments \citep{Schaye-2015,Crain-2015}, reproduces many key observational results such as the specific star formation
rates \citep{2015-Furlong}, the passive galaxy fraction \citep{2015-Trayford}, the Tully-Fisher relation \citep{Schaye-2015}, 
the H$_{2}$ mass functions at $z\,$=$\,0$ and several atomic and molecular gas scaling relations \citep{2015-Lagos, 2016-Bahe, 2017Crain}.
The fact that we can detect ring galaxies in a statistical sample of simulated galaxies implies that
we can use these systems as a higher-order constraint on the models because morphology was not used to tune the parameters in EAGLE
(see \citet{Crain-2015} for more details). Thus, our study represents a true prediction of the simulation and by comparing
these results with the observations we hope to learn whether the numerical treatments of the ISM, star formation, and feedback 
are adequate enough to reproduce a realistic ring-morphology population. 
We also  study the general characteristics of collisional ring galaxies in the EAGLE simulations including their star formation
rates (SFRs), colours, metallicities, atomic (HI) and molecular hydrogen (H$_{2}$) gas scaling relations.
We aim to provide a thorough description of the ISM properties of collisional ring galaxies in the EAGLE simulations and the possible reasons
leading to their high HI masses or molecular gas deficiencies.\\

This paper is organised as follows: In Section 2, we briefly describe the key characteristics of the simulation, the
subgrid physics included in EAGLE and the mock optical images generated by the SKIRT code \citep[][hereafter T17]{2017Trayford}.
Section 3 describes our selection criteria  applied to select ring galaxies in the simulation.
In section 4, we present the characteristics of ring galaxies in EAGLE, including their number density, halo masses and the
concentrations of the haloes containing rings, colour-magnitude diagram, HI-stellar mass relation as well as the SFR-stellar mass
of ring galaxies, at different redshifts. We also present a more detailed study of the ISM properties in ring galaxies. In Section 5 we present
the origin and the formation history of rings in the EAGLE simulation. We present our discussion and conclusion in Section 6.
For all the calculations, we adopt the cosmology used for the EAGLE simulation, a $\Lambda$CDM cosmology with $\Omega_{m}$=$0.307$,
$\Omega_{\Lambda}$=$0.693$, $\Omega_{b}$=$0.048$, $\sigma_{8}$=$0.8288$, and $H_{0}\,$=$\,67.77\,$km\,s$^{-1}$~Mpc$^{-1}$,
consistent with  Planck measurements \citep{2014Planck}.\\

\section{SIMULATED GALAXY SAMPLES}
\subsection{Overview of the EAGLE Simulation}
The EAGLE project is a suite of hydrodynamical simulations designed to explore the evolution of the Universe's 
baryonic matter (gas, stars and  massive  black  holes) and  dark  matter from  a redshift of $z=127$  until  $z=0$ \citep{Schaye-2015,Crain-2015}.   
The simulations were constructed assuming the standard cosmological paradigm (the so called $\Lambda$CDM 
cosmological model) and run with an extensively modified version of the N-body TreePM smoothed particle hydrodynamics 
(SPH) code GADGET 3 \citep{Springel-2005,2008Springel}. The main updates to the standard GADGET 3 code feature modifications 
to the hydrodynamics algorithm, and the incorporation of subgrid modules that capture the unresolved physics acting on
scales below the resolution limit of the simulations. The modified SPH algorithm, referred to as `Anarchy',
includes the implementation of the pressure-entropy formulation of SPH \citep{2013Hopkins}, the artificial viscosity switch 
proposed by \citet{2010Cullen}, the artificial conduction switch similar to that proposed by \citet{2008Price}, the C$_2$
smoothing kernel of \citet{Wendland}, and the timestep limiter from \citet{2012Durier}. The impact of the above modifications to
the standard GADGET 3 code and its effect on the simulated galaxies is described in \citet{2015Schaller}.\\

The EAGLE simulations incorporate state-of-the-art subgrid physics based on those used for the OWLS \citep{2010Schaye} and the
GIMIC \citep{2009Crain} projects. These subgrid physics models include element-by-element radiative cooling and photoheating 
rates \citep{2009WiersmaSchayeSmith}, star formation as a pressure law \citep{2008Schaye} and a metallicity-dependent
density threshold \citep{2004Schaye}, stellar evolution and element-by-element chemical enrichment \citep{2009Wiersma},
stellar feedback as energy injected from core-collapse supernovae \citep{2012DallaVecchia}, as well as accreting black holes (BH)
and AGN feedback \citep{2015osas-Guevara}. The efficiency of the stellar feedback and the BH accretion
were calibrated to match the observations of the  galaxy stellar mass function (GSMF) at $z = 0.1$, while the AGN feedback was 
calibrated to match the observed relation between stellar mass and BH mass. The EAGLE simulation has had unprecedented 
successes as it reproduces many key observational datasets (that were not considered during the calibration) including 
the stellar mass function of galaxies \citep{2015-Furlong}, the stellar mass-size relation \citep{2017-Furlong}, the colour
distribution of galaxies \citep{2015-Trayford}, the cold gas contents of galaxies throughout cosmic time 
\citep{2015-Lagos, 2016Lagos, 2016-Bahe, 2017Crain}, and the evolution of the star formation rate with redshift \citep{2017Katsianis},
among others.\\

\begin{table}
\centering
\caption{Key parameters of the EAGLE Ref-L100N1504 simulation used in this paper. The EAGLE simulation adopts
a softening length of $2.66\,$ckpc at $z\geq2.8$, and $0.7\,$pkpc at $z < 2.8$.}
\label{my-label}
\begin{tabular}{ll}\hline \hline
Comoving box size& $100\,$cMpc \\
Number of particles &  $2\times1504^3$\\
Gas particle mass & $1.81\times10^6\,$M$_{\odot}$ \\
Dark matter particle mass& $9.7\times10^6\,$M$_{\odot}$  \\
Softening length& $2.66\,$ckpc \\
Max. gravitational softening length& $0.7\,$pkpc   \\ \hline
\end{tabular}
\end{table}

For this work, we use the largest reference model Ref-L100N1504 (hereafter Ref-100), which is a cubic cosmological volume of $100$ comoving
megaparsecs (cMpc) on a side, with dark matter particle mass of $9.7\times10^{6}\,$M$_{\odot}$, initial gas particle mass
of $1.81\times10^6\,$M$_{\odot}$, and equal numbers of baryonic and dark matter particles ($1504^3$). The Plummer equivalent 
gravitational softening lengths are set to 1/25 of the initial mean interparticle spacing and are  $0.7\,$ proper kiloparsecs
(pkpc) at redshifts $z < 2.8$, and $2.66\,$comoving kiloparsecs (ckpc) at earlier times, which means that the Jeans scales
in the warm ISM are marginally resolved. The volume of Ref-L100N1504 provides a wide range of galaxy morphologies in a sample of $\sim30,000\,$galaxies resolved by $>\,1,000$ 
star particles and $\sim3,000\,$galaxies resolved by $>\,10,000\,$ star particles at redshift $z=0.1$. Table 1 presents a summary of the key features of the reference model
Ref-L100N1504. The properties of the particles in the simulation were recorded for $29$ snapshots between redshifts $20$ and $0$,
which translates to time span range between snapshots of $0.3-1\,$Gyr. However, finer time resolution are also available, referred to as `snipshots' in
\citet{2017Crain}, in which a smaller set of particle properties are saved at $400$ redshifts between $20\,\leq\,z\,\leq0$.
It is important to note that only $200$ of the $400$ snipshots were used to construct the merger trees; 
the merger trees are computationally expensive to construct. Hence, the time resolution of the snipshots span between $0.05-0.3\,$Gyr. We use the merger trees available in the EAGLE 
database \citep{2016McAlpine} to trace the evolution of ring galaxies in the snapshots of the simulation, while for snipshots
we use the private merger trees catalogue. All these merger trees were constructed as in \citet{Qu2017}.\\

\subsection{Mock observations \& Images}

\begin{figure*}
\centering
\includegraphics[width=1.02\columnwidth]{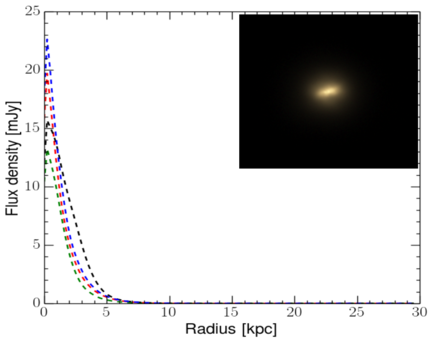}\includegraphics[width=1.02\columnwidth]{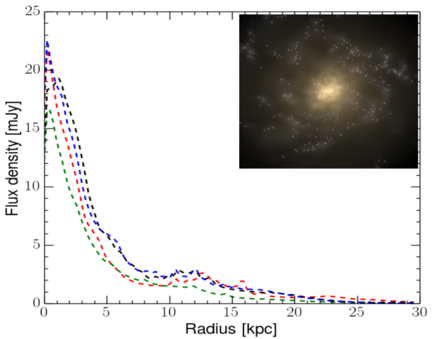}\\
\includegraphics[width=1.025\columnwidth]{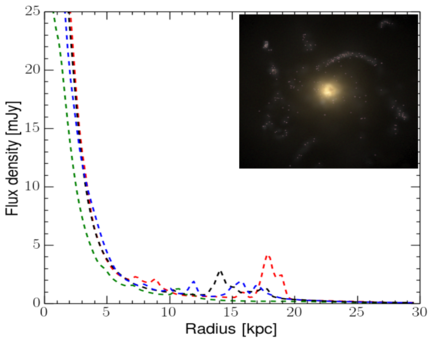}\includegraphics[width=1.02\columnwidth]{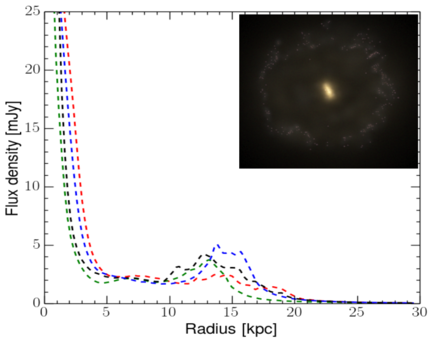}
\caption{The kernel smoothed radial flux profile in four different strips (shown by the coloured dashed-lines) of the face-on $u$-band image of an elliptical (top left), spiral 
(top right), and two morphologically disturbed (interacting; bottom panels) galaxies at $z = 0$. The radial profiles 
describe the flux distribution of that respected galaxy from its centre to its outer edge along its major and minor axes. The inset shows the synthetic $gri$ composite  
image of these galaxies. Only one of the two interacting galaxies is classified as a ring galaxy (bottom right).}
\label{3b}
\end{figure*}

\citet[][hereafter T17]{2016Camps, 2017Trayford} present a novel method to generate mock synthetic optical images of galaxies in
the EAGLE simulations that includes the effects of dust using the SKIRT Monte Carlo radiative transfer code 
\citep{2003Baes,2011Baes,2015Camps}. T17 compute a full spectral energy distribution (SED) for each star particle 
using the the GALAXEV population synthesis models of \citet{2003Bruzual}, taking into account the stellar ages, smoothed 
metallicities and initial masses of the star particles. They also use the MAPPINGS photoionisation code \citep{2008Groves}
to describe the effects of dust associated with star-forming regions, nebular line and continuum spectra throughout
the HII regions. The SKIRT Monte Carlo code then uses these sets of sources and dust distributions to determine 
the path of the monochromatic photons (absorption \& scattering) until they hit the user specified detector.\\

T17 construct images in $ugriz$ SDSS bands \citep{2010Doi} of $3,624$ galaxies with M$_{*}> 10^{10}$M$_{\odot}$ 
in the EAGLE hydrodynamical simulation at redshift $z=0.1$. These observations are available in three orientations: edge on, face
on and randomly orientated with respect to the galactic plane. Each image is $256\times256$ pixels in size, with field of view (FOV)
of $60\,$kpc, and with the detector camera placed $50\,$Mpc away from the galaxy centre. The mock observations are performed
as if each galaxy is in isolation, hence disregard all the contribution from other sources and structures along the line of sight or
closer than $60\,$kpc in projection.\\

\section{Selection of Ring Galaxies in EAGLE}

\begin{figure*}
\centering
\includegraphics[width=2.05\columnwidth]{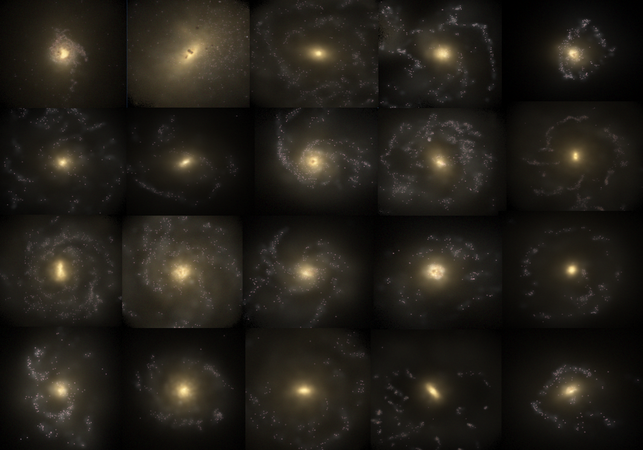}
\caption{Three-colour $gri\,$ mock images of a subsample of the interacting galaxy candidates selected by the algorithm at $z$=$0$. 
The majority of these galaxies are morphological disturbed or interacting systems and a small fraction of these systems 
are real ring galaxies (far right column). These images are $60\,$kpc on a side and are publicly available from 
the EAGLE database \citep{2016McAlpine}.}
\label{3c}
\end{figure*}

We develop a Python algorithm specifically designed to identify ring galaxies in the Ref-100 simulation box.
The routine makes use of the face-on orientation $ugriz$-images created by T17 and
broadly quantifies galaxy morphologies into ellipticals and lenticulars (early-type), spirals (late-type), and morphologically disturbed (interacting) systems.
This algorithm relies on the radial flux profiles of these galaxies and works as following:

\begin{figure}
\centering
\includegraphics[width=1.05\columnwidth]{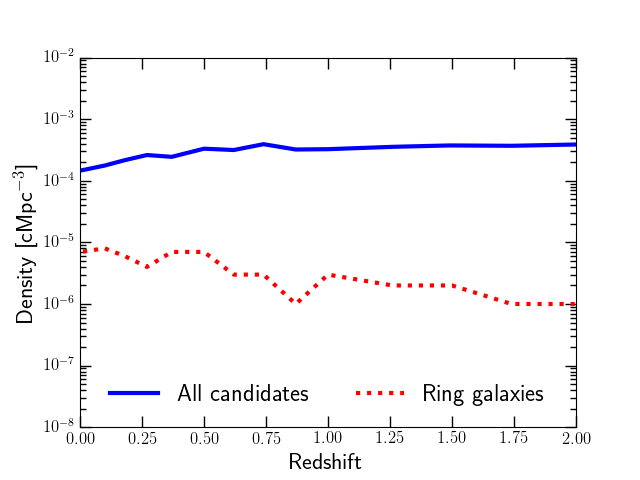}
\caption{The number density of the interacting or disturbed candidates against redshift (blue line) and the
visually selected subsample of ring or C-shape morphology candidates (red dotted line). The number of interacting galaxy candidates 
increases with redshift in agreement with the interactions/mergers rates (refer to Figure \ref{4a}).}
\label{3d}
\end{figure}

\begin{itemize}
 \item Each galaxy is divided into four strips, each  $4\,$kpc in width, along the major and the minor axis of the galaxy.
For each strip, we construct a radial flux profile using the rest-frame face-on $u$-band images, and smooth this profile using 
a kernel function with a Gaussian scale. By definition each galaxy is placed at the centre of image, hence the radial profile describes the 
flux distribution of each respective galaxy from its centre to its outer edge.
For early-type (elliptical and lenticular) galaxies, the flux distribution varies rapidly with  radius
with a high peak at the centre of the galaxy that decreases abruptly with radius. The case is slightly different for late-type
(spiral) galaxies, where the flux decreases gradually with radius and the spiral arms produce flux increments (bumps) throughout the 
radial flux profile. The radial profile of interacting (morphologically disturbed) galaxies is similar to that of the 
late-type galaxies, albeit the bumps are more extended and the flux increments are higher.
It is important to note that no changes to the final results will occur if the width of the strip is slightly
larger than $4\,$kpc.\\

\begin{figure*}
\centering
\includegraphics[width=2.0\columnwidth]{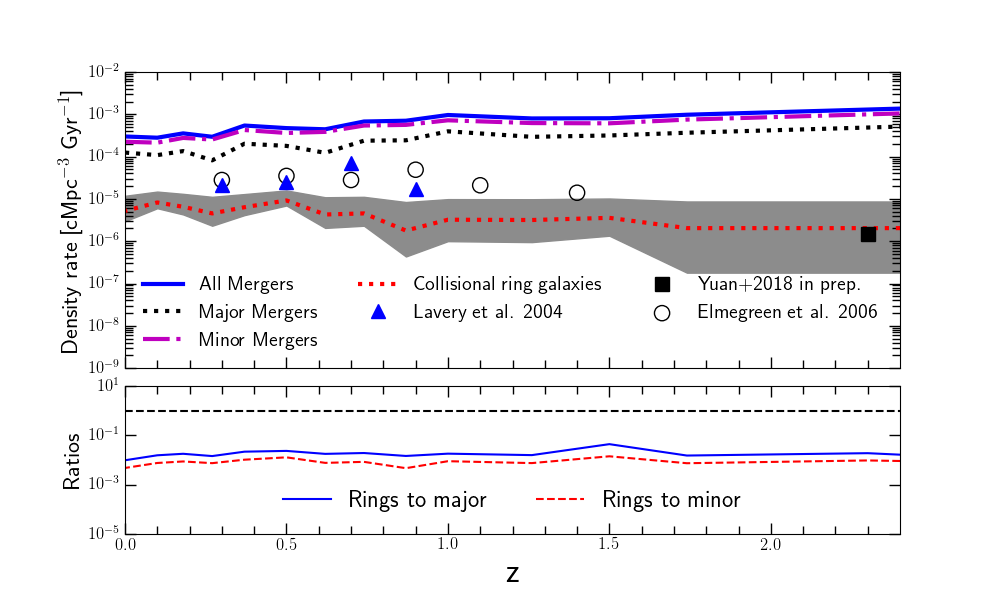}
\caption{Upper Panel: The formation density rate of ring galaxies in the EAGLE simulations (red dotted line) compared to the number density from
observations (black unfilled circles from \citealt{Elmegreen-2006}, blue triangles from \citealt{Lavery-2004}, and filled square from Yuan et al. 2018 in prep.). The grey shaded region marks
the Poisson errors in the EAGLE number density.
The blue line shows the number density of all mergers in the simulation with a stellar mass ratio $\gtrsim\,0.1$, while the black dotted 
and magenta dash-dotted lines show the contribution from major (mass ratios $\gtrsim\,0.3$) and minor mergers (mass ratios between $0.1\,-\,0.3$), respectively.
EAGLE agrees very well  with the ring galaxy observations. Lower Panel: The ratio between the collisional ring galaxies number density to the major (blue line) and minor 
(red dashed line) merger rates. The black dashed line represents equality.}
\label{4a}
\end{figure*}

\item  Next the algorithm searches and quantifies the flux increments (bumps) in the radial profile of each galaxy
and uses both the extent (width) and the amount of flux boost in each bump as a broad indicator to quantify the galaxy morphology.
Early-type galaxies are the simplest to segregate, in which  there are no flux increments with radius in the profiles and the 
flux decreases smoothly with increasing radius. To distinguish between late-type and morphologically disturbed systems, we
use the width of the bump and the flux increment present in the galaxy radial profile. Galaxies with continuous bump(s) that
extends for $\gtrsim2\,$kpc in at least two strips and the flux boost in each bump is $\geq10\%$ are considered as 
interacting candidates. Figure \ref{3b} presents an example of the kernel smoothed radial flux distribution in four strips of the face-on image of an
elliptical, spiral, and two morphologically disturbed (interacting) galaxies at redshift $z=0$, respectively.
The inset shows the $u$-band image of the same galaxies. The flux distribution in the elliptical galaxy has no bumps and 
decreases continuously with increasing radius. On the other hand, the radial flux profile in the spiral and the interacting
candidates shows many bumps with increasing radius. However, the difference between the two is that the spiral 
galaxy shows less boost in the flux at each bump ($<10\%$ ) compared with the two interacting systems in the bottom panel.\\

\item Lastly, the algorithm makes a list of all the interacting candidates and creates plots similar to Figure \ref{3b} 
for those galaxies. The final step is to visually inspect the interacting galaxies subsample and only select
galaxies that have a ring or C-shape morphology. Visual inspection is necessary in order to distinguish the P-type (collisional)
and resonant O-type ring galaxies and to determine the presence of any weak bar-like or spiral structure.
Figure \ref{3c} shows a subsample of the $gri\,$ mock images of the interacting
galaxy candidates selected by the algorithm at $z$=$0$, the majority of these galaxies are morphologically disturbed or interacting 
systems and a smaller fraction are the real ring galaxies. In this figure the true ring galaxies are located at the far right column. 
We visually inspect all the interacting galaxy candidates generated by the algorithm
between redshift $z=0$ to $z=2$ and select only the candidates with a ring or C-shape morphology. Figure \ref{3d} shows
the number density of the interacting or disturbed candidates as a function of redshift (blue line) along with the
visually selected subsample of ring or C-shape morphology candidates (red dotted line). The number of disturbed or potentially
interacting galaxies increases with redshift, which is consistent with other studies \citep[e.g.,][]{Abraham1996, Fakhouri-2010,Bluck-2012, Man-2016}. The total number 
of interacting candidates varies between $150$ at redshift $z=0$ up to $350$ candidates at $z=2$.\\
\end{itemize}
To investigate the success rate of our algorithm, we search for ring galaxies visually at redshifts $z=0, 0.5$ and $1.0$ using 
the $gri$ composite images available in the database \citep{2016McAlpine}. We inspect all the galaxies in these three snapshots that have a stellar mass of 
M$_{*}\,\geq\,10^{10}$M$_{\odot}\,$(corresponding to the typical masses of observed ring galaxies) 
and record all of the ring galaxies in these snapshots. Then, we run this semi-automated algorithm and 
compare the visually inspected ring galaxies with the the list of interacting candidates that results at redshifts $z=0, 0.5$ and $1.0$.
We find that the list of interacting candidates generated by the semi-automated algorithm recovers all the ring galaxies
that were visually found. Further, to test the reliability of the algorithm, we set the 
extent of the bump to $3\,$kpc and $4$kpc instead of $2\,$kpc. The number of candidates in the former two cases are similar to the number of candidates that results from 
setting the required extent to $2\,$kpc. However, if the  extent of the bump decreases to $1\,$kpc or less, the number 
of candidates increases by up to $20\%$ more at a given redshift. The majority of the new candidates are spiral 
galaxies, with no real ring galaxies. Hence, setting the width of the bump to $2\,$kpc is optimal and reduces the number 
of false candidates appearing as a result of the spiral arms in late type galaxies. 
Further, we apply our algorithm to the EAGLE higher resolution smaller box Ref-L025N0752 \citep{Schaye-2015} and search
for ring galaxies between redshifts $z\,=\,0\,$ to $z\,=\,2$, 
with the aim of uncovering any obvious resolution effects on the formation process of ring galaxies.
We find no ring galaxies in this simulation box, which agrees with the expected number density of ring galaxies from the observations. In Ref-L025N0752 box 
($25^{3}$ cMpc$^{3}$), the expected number of ring galaxies in each snapshot is $< 0.06$, using the number density of $4.3\times10^{-6}\,$Mpc$^{-3}$ from \citet{Lavery-2004}.
This result provides no evidence to suggest that the ring galaxies detected in Ref-100 are due to resolution effects.\\

\begin{figure}
\centering
\includegraphics[width=1.0\columnwidth]{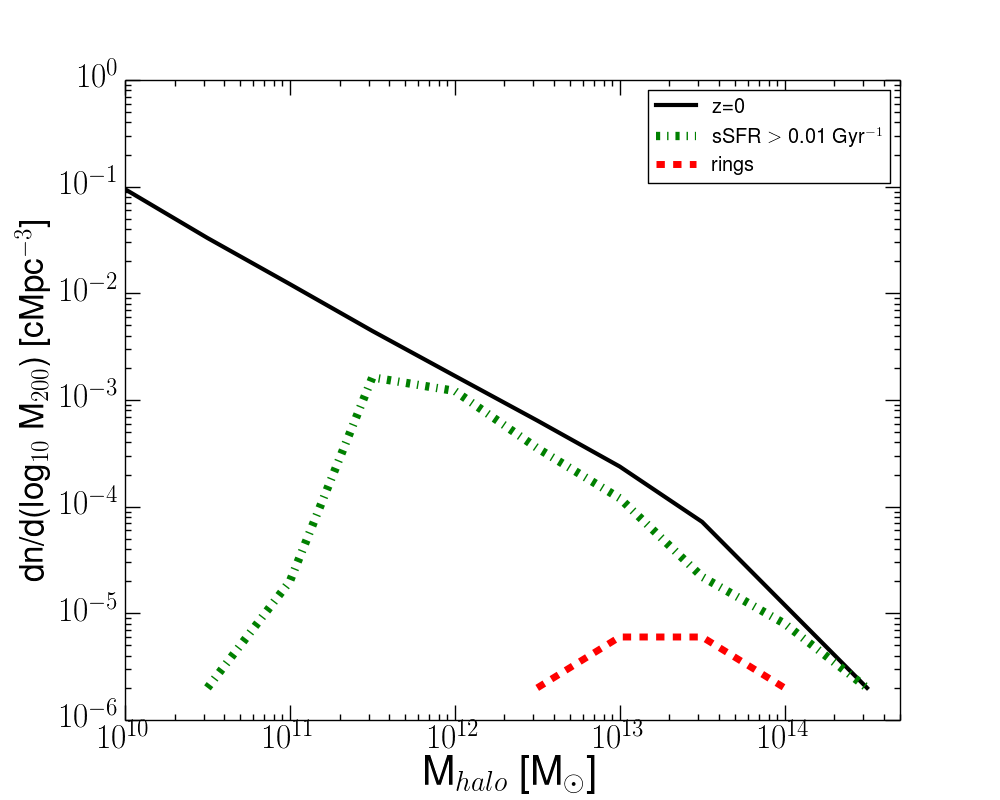}
\includegraphics[width=1.0\columnwidth]{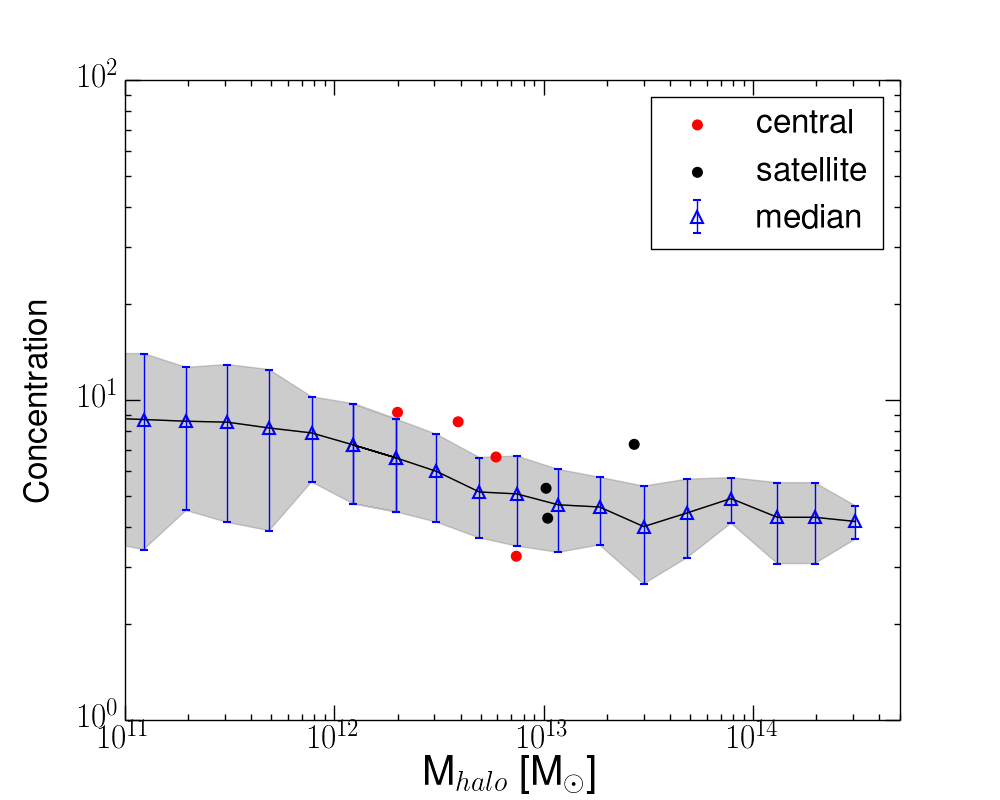}
\caption{Upper Panel: The halo mass function for all the haloes at $z$=$0$ (black line), and for the subsample of haloes hosting star-forming central galaxies with sSFR$\,>0.01\,$Gyr$^{-1}$ 
(green dash-dotted line) and ring galaxies (red dashed line). Lower Panel: The concentration of the haloes as a function of their mass at $z$=$0$. 
The grey shaded region shows the $68^{\x{th}}$ percentile while the median is shown as blue triangles. The red and black symbols are the concentrations of haloes hosting  central and non-central ring galaxies, respectively.}
\label{4b}
\end{figure}

To distinguish between the P and O-type ring galaxies, we crossmatch our sample  with the sample of barred galaxies in EAGLE 
reported in \citet[][hereafter A17]{Algorry-2017}. A17 analyse the central galaxies in EAGLE that lie within mass range of
$10.6<\x{log}_{10}(\x{M}_{*}/\x{M}_{\odot})<11$ at redshift $z\,=\,0$ and classify galaxies via the amplitude of the
normalised $m\,=\,2$ Fourier mode of the azimuthal distribution of their disk particles.
Galaxies with a normalised amplitude (A$_{2}^{\x{max}}$) $< 0.2$ are unbarred systems, those with an amplitude in the range of
$0.2 < \x{A}_{2}^{\x{max}} < 0.4$ have a weak bar, while strongly barred galaxies have an amplitude A$_{2}^{\x{max}}$ $> 0.4$.
At redshift $z\,=\,0$, we identify seven systems as ring galaxies, one of which has a strong bar 
(A$_{2}^{\x{max}}$ $> 0.4$) and the remainder are unbarred galaxies. At higher redshift, we explore the formation history of our
ring galaxy sample to distinguish between barred and collisional ring galaxies. This is discussed in Section 5.\\

\begin{figure*}
\centering
\includegraphics[width=0.34\columnwidth]{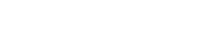}\includegraphics[width=0.34\columnwidth]{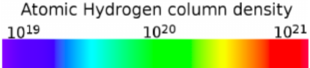}\includegraphics[width=0.34\columnwidth]{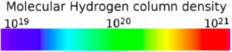}
\includegraphics[width=0.34\columnwidth]{whit.png}\includegraphics[width=0.34\columnwidth]{hydrogen.png}\includegraphics[width=0.34\columnwidth]{molecularhydrogen.png}\\
\includegraphics[width=0.3365\columnwidth]{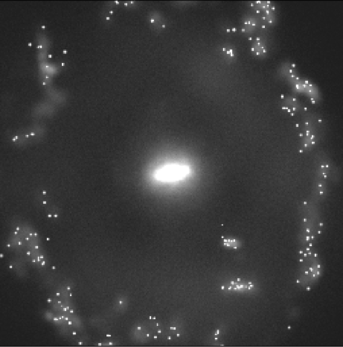}\includegraphics[width=0.34\columnwidth]{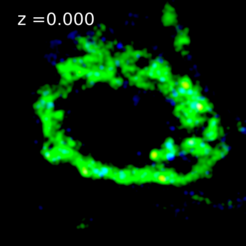}\includegraphics[width=0.34\columnwidth]{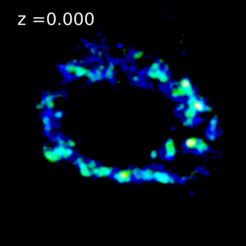}
\includegraphics[width=0.35\columnwidth]{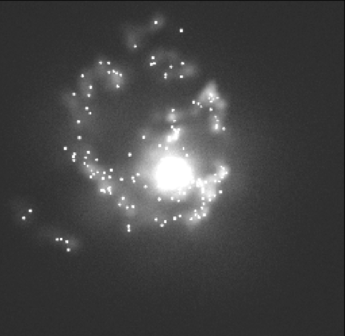}\includegraphics[width=0.34\columnwidth]{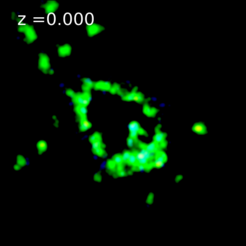}\includegraphics[width=0.34\columnwidth]{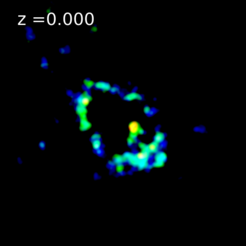}\\
\includegraphics[width=0.34\columnwidth]{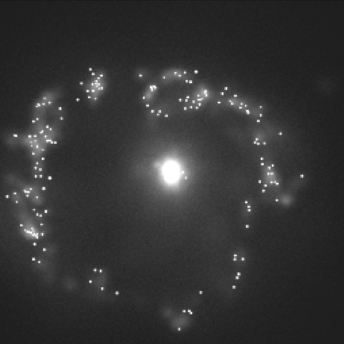}\includegraphics[width=0.34\columnwidth]{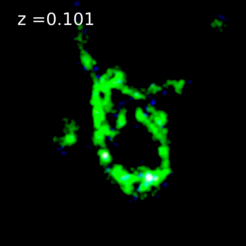}\includegraphics[width=0.34\columnwidth]{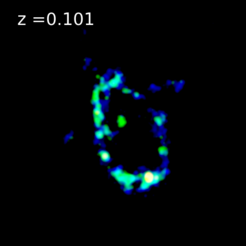}
\includegraphics[width=0.34\columnwidth]{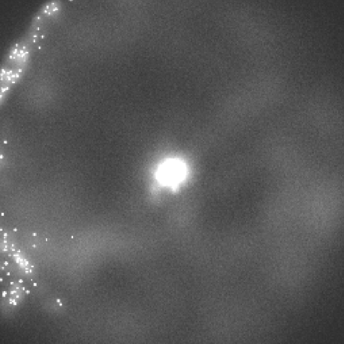}\includegraphics[width=0.34\columnwidth]{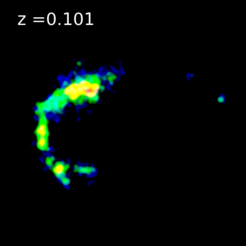}\includegraphics[width=0.34\columnwidth]{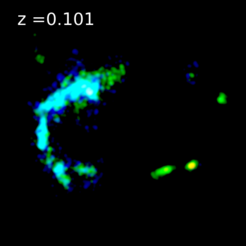}\\
\includegraphics[width=0.34\columnwidth]{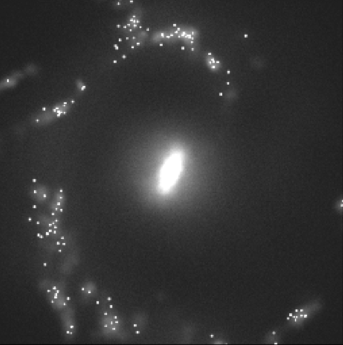}\includegraphics[width=0.34\columnwidth]{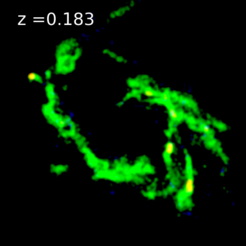}\includegraphics[width=0.34\columnwidth]{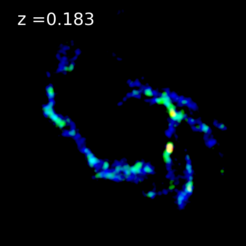}
\includegraphics[width=0.34\columnwidth]{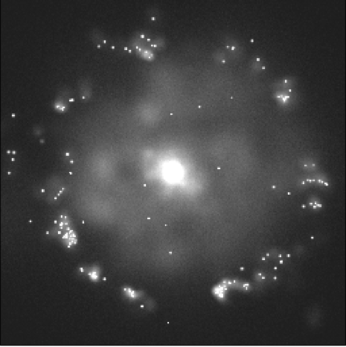}\includegraphics[width=0.34\columnwidth]{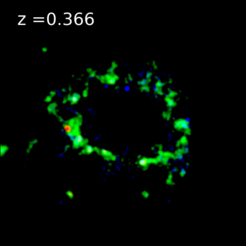}\includegraphics[width=0.34\columnwidth]{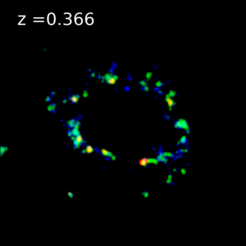}\\
\includegraphics[width=0.34\columnwidth]{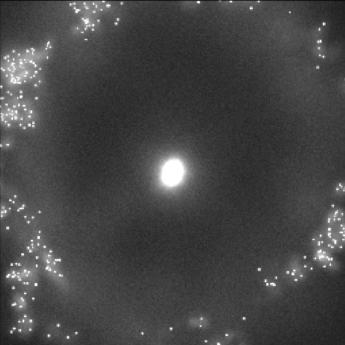}\includegraphics[width=0.34\columnwidth]{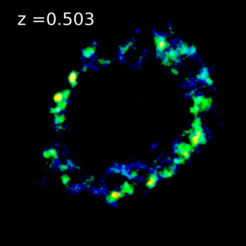}\includegraphics[width=0.34\columnwidth]{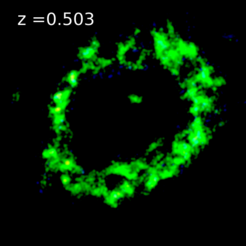}
\includegraphics[width=0.34\columnwidth]{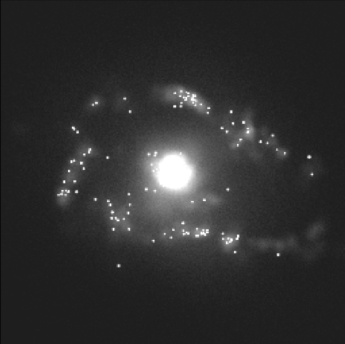}\includegraphics[width=0.34\columnwidth]{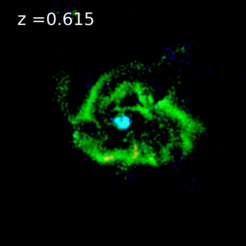}\includegraphics[width=0.34\columnwidth]{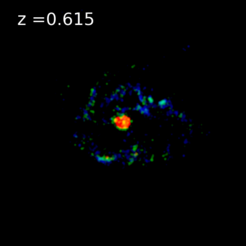}\\
\includegraphics[width=0.34\columnwidth]{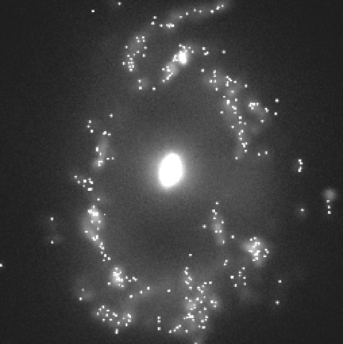}\includegraphics[width=0.34\columnwidth]{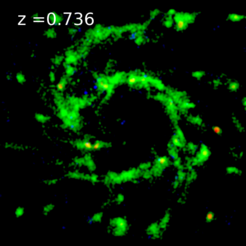}\includegraphics[width=0.34\columnwidth]{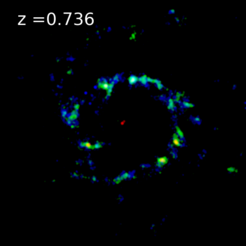}
\includegraphics[width=0.34\columnwidth]{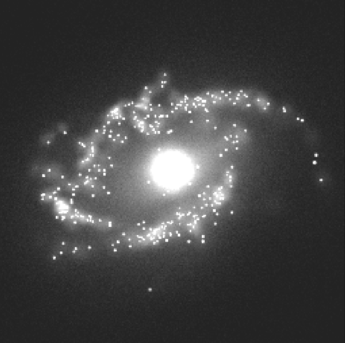}\includegraphics[width=0.34\columnwidth]{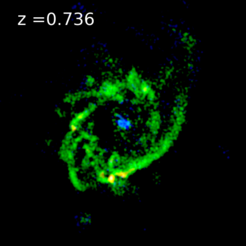}\includegraphics[width=0.34\columnwidth]{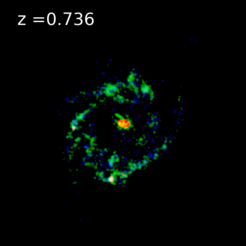}
\includegraphics[width=0.34\columnwidth]{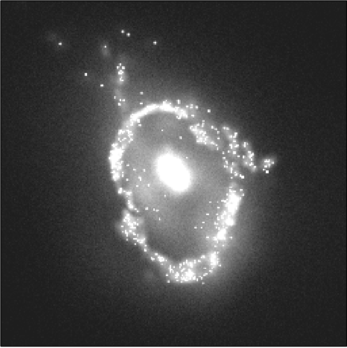}\includegraphics[width=0.34\columnwidth]{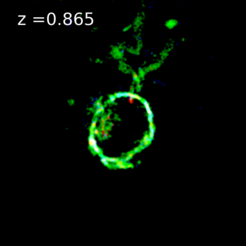}\includegraphics[width=0.34\columnwidth]{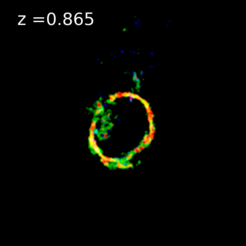}
\includegraphics[width=0.34\columnwidth]{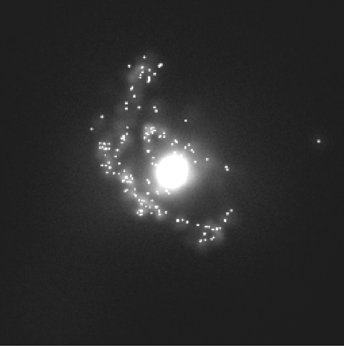}\includegraphics[width=0.34\columnwidth]{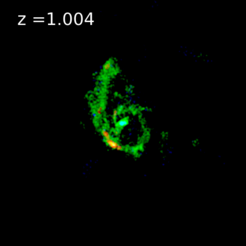}\includegraphics[width=0.34\columnwidth]{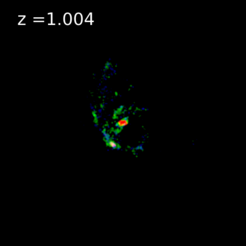}\\
\includegraphics[width=0.34\columnwidth]{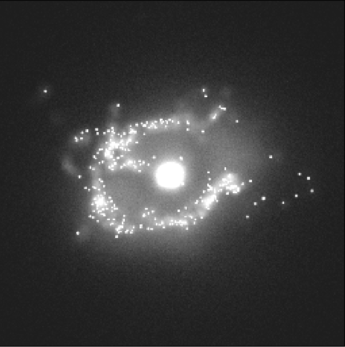}\includegraphics[width=0.34\columnwidth]{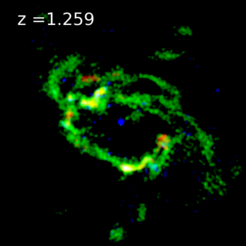}\includegraphics[width=0.34\columnwidth]{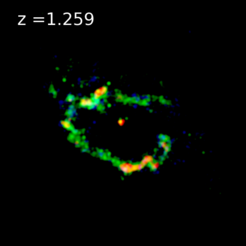}
\includegraphics[width=0.34\columnwidth]{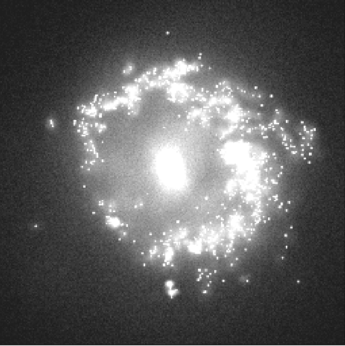}\includegraphics[width=0.34\columnwidth]{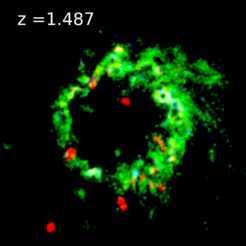}\includegraphics[width=0.34\columnwidth]{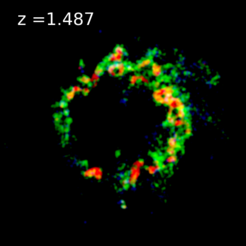}\\
\caption {Visualisation of fourteen randomly selected collisional ring galaxies at a redshift range between $z\,$=$\,0$ and $1.5$ identified
in EAGLE. Each row corresponds to the face-on $u$-band image (left), the HI (middle) and H$_2$ (left) column density maps of
the same galaxy. The colours in the column density images are coded according to the colour bar at the top row and
  are in units of cm$^{-2}$. The $u$-band images are $60\,$kpc on a side while the HI and H$_2$ column density maps
are $100\,$kpc on a side.}
\label{4c}
\end{figure*}

\section{Characterisation of the Ring Galaxy Population in EAGLE}
In this section, we study the number density, environment and properties of ring galaxies. We pay special attention to the relation 
between the gas abundance and star formation in the ring galaxies to understand the origin of their star formation deficiency.
\subsection{Rings as Tracers of Galaxy Mergers}
\noindent The upper panel of Figure \ref{4a} shows the formation density rate of collisional ring galaxies identified in the EAGLE 
simulations (red dash-line) and in observations as a function of redshifts. The grey shaded region marks the one-sigma scatter 
in the EAGLE number density rate\footnote{The number density rate is the number of collisional ring galaxies per comoving volume per time interval.
Here, we adopt the time interval between each snapshot for our calculations.}, refer to Table \ref{my-labele} for the number of ring galaxies identified in EAGLE at each snapshot. The blue triangles show the number density rate of these galaxies identified (visually)
in $162$ deep Hubble Space Telescope (HST) Wide Field Planetary Camera 2 archival images and reported in \citet{Lavery-2004}.
The black unfilled circles show the number density rate measured in \citet{Elmegreen-2006} by visually inspecting the deep archival images
of the Galaxy Evolution from Morphology and SEDs survey \citep[GEMS;][]{2004Rix} and the Great Observatories 
Origins Deep Survey \citep[GOODS;][]{2004Giavalisco}.
The filled square presents a new measurement for the number density of ring galaxies in the redshift range $1.8 < z < 2.8$ 
identified visually in the COSMOS field of the FourStar Galaxy Evolution (ZFOURGE) catalogue field images \citep{2016Straatman}, which will be reported in Yuan et al 2018 (in prep.).
Galaxy merger rates in the EAGLE simulation, as presented in \citet{2018Lagos}, are also presented in this figure for comparison. 
The major merger rates (stellar mass ratios of $\geq 0.3$) are shown as the black dotted line, minor merger rates (stellar mass ratios of $0.1-0.3$) as
the magenta dash-dotted line and all merger rates (minor+major) as the blue line. \\

The lower panel of  Figure \ref{4a} shows the ratio between the collisional ring galaxies number density and the major 
(blue line) and minor (red dashed line) merger rates. The black dashed line represents  equality. This figure 
shows that collisional ring galaxies are biased tracers of merger rates and that this bias is independent of redshift.
Also, the number density of collisional ring galaxies found in EAGLE simulations broadly agrees with the observed number density within the uncertainties.
This is a success for the EAGLE project especially because the calibration of the subgrid physics in these simulations did not include galaxy morphology \citep{Crain-2015}.
However, it is important to point out the moderate difference in the theoretical prediction of the number density value at $z\sim0.7$ and $0.9$;
this difference could be due to the various systematic uncertainty sources in the observed number density such as cosmic variance and the uncertainty in the 
redshift measurements. For instance, the majority of the observed ring galaxies in \citet{Lavery-2004} have indirect redshift measurements using a `standard' 
absolute $V$-band magnitude for collisional ring galaxies \citep{1997Appleton} to constrain the redshift interval of their  sample. \\

\subsection{The Host Halos of Eagle Ring Galaxies}
\noindent The upper panel of Figure \ref{4b} shows the mass function for all the haloes at redshift $z\,$=$\,0$ (black line), 
for haloes hosting a star-forming central galaxy (green dash-dotted line) and for haloes hosting a ring galaxy (red dashed line) at $z\,$=$\,0$. Star-forming 
central galaxies are those with specific star formation rates (sSFR$\,$=$\,$SFR/M$_{*}$) $\,>0.01\,$Gyr$^{-1}$ \citep{2015-Furlong}. The lower panel of Figure 
\ref{4b} shows the concentration of haloes (the ratio between the virial radius to the characteristic radius of that halo) as a function of their mass
at $z$=$0$. The grey shaded region is the one sigma deviation from the median value (blue triangles). The red and black symbols show  the concentrations of haloes hosting
central and non-central ring galaxies, respectively. This figure shows that ring galaxies found in EAGLE live in 
massive groups that are preferentially more concentrated than other groups that host no ring galaxies at 
fixed halo mass, in spite of the low number statistics. This is broadly consistent with the observations, which show that ring galaxies are located within galaxy
groups that have at least one companion galaxy \citep{Romano2008}. It is  reasonable to think that ring galaxies form 
in dense environments such as compact groups, since a common formation mechanism is based on collisions and drop-through interactions. 
However, ring galaxies are not found in galaxy clusters, even though the galaxy population density is higher in the
cluster environment. The main reason for this is that the high velocity dispersions and/or any off centre encounters 
between the ring galaxy and other cluster members will disrupt the ring more quickly than in the group environment. 
This explains the decline in the halo mass function of ring galaxies (red line) with masses larger than 
$10^{13.3}\,$M$_{\odot}$.\\

\begin{figure}
\centering
\includegraphics[width=0.48\columnwidth]{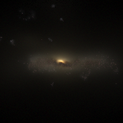}\includegraphics[width=0.48\columnwidth]{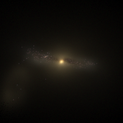}\\
\includegraphics[width=0.48\columnwidth]{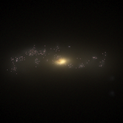}\includegraphics[width=0.48\columnwidth]{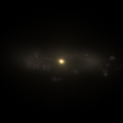}\\
\includegraphics[width=0.48\columnwidth]{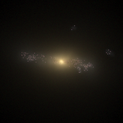}\includegraphics[width=0.48\columnwidth]{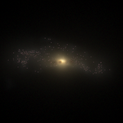}
\caption{Three-colour $gri$ edge-on images of a subsample of EAGLE ring galaxies identified in  the redshift range between $z=0$ and $1$. These images are $60\,$kpc on 
a side and are available in the EAGLE database webpage \citep{2016McAlpine}.}
\label{fig_edge-on}
\end{figure}

\begin{figure}
\centering
\includegraphics[width=0.48\columnwidth]{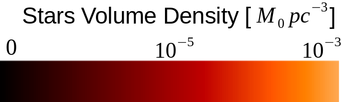}\includegraphics[width=0.48\columnwidth]{whit.png}\\
\includegraphics[width=0.48\columnwidth]{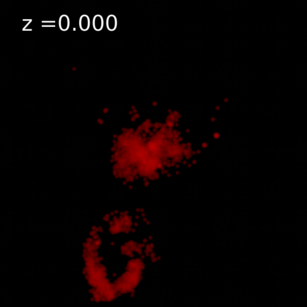}\includegraphics[width=0.48\columnwidth]{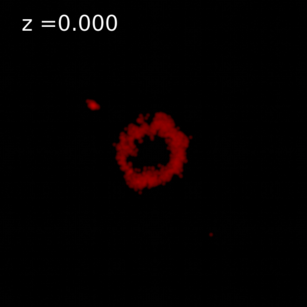}\\
\includegraphics[width=0.48\columnwidth]{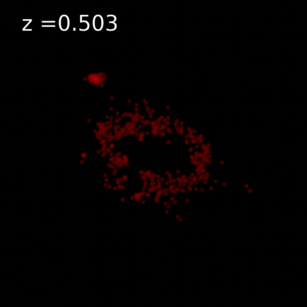}\includegraphics[width=0.48\columnwidth]{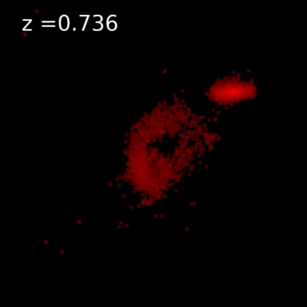}\\
\includegraphics[width=0.48\columnwidth]{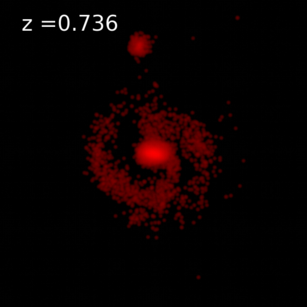}\includegraphics[width=0.48\columnwidth]{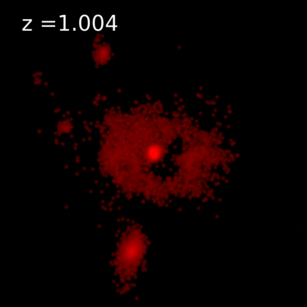}
\caption{Visualisation of the stellar volume density view in six randomly selected collisional ring galaxies identified
in EAGLE between $z=0$ and $1$, as labelled in each panel. In each view, a second galaxy (satellite) appears within a close distance (roughly one ring diameter) from the 
central ring galaxy. These maps are $200\,$kpc on a side.}
\label{fig_starview}
\end{figure}

\begin{figure*}
\centering
\includegraphics[width=0.84\columnwidth]{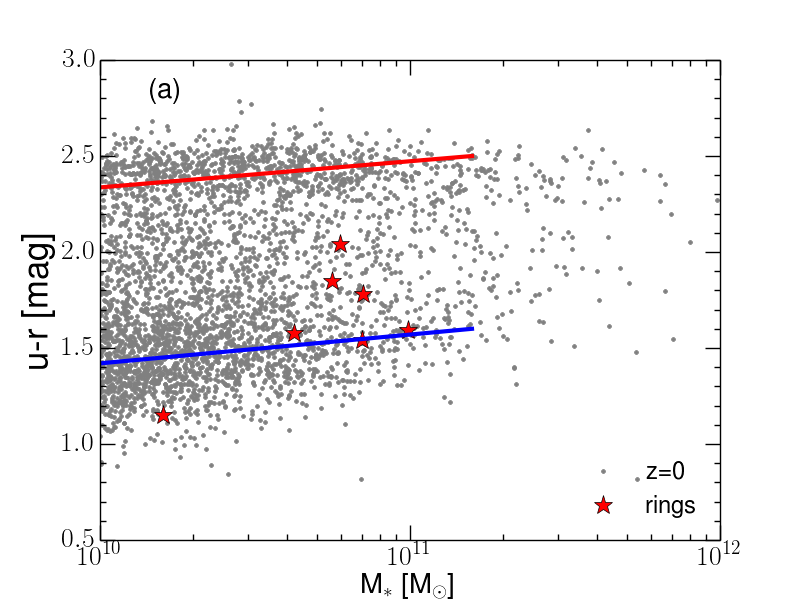}\includegraphics[width=0.84\columnwidth]{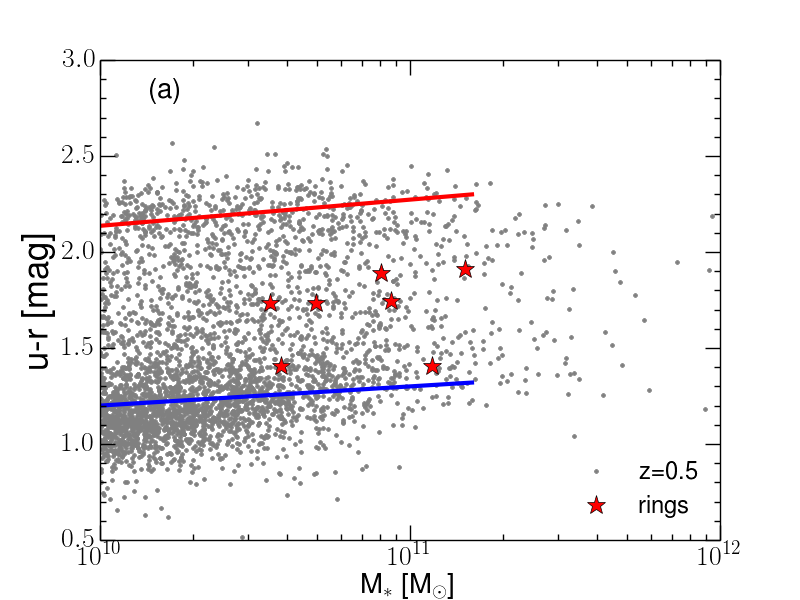}\\
\includegraphics[width=0.84\columnwidth]{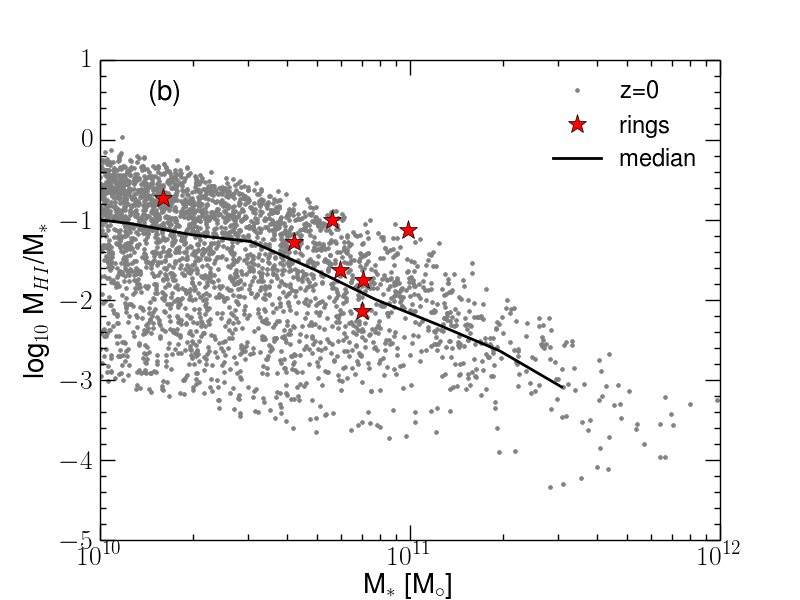}\includegraphics[width=0.84\columnwidth]{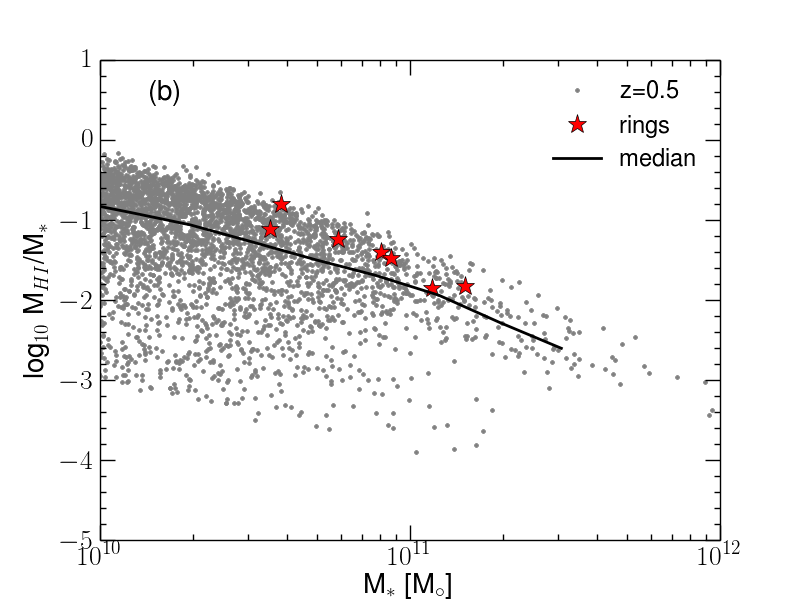}\\
\includegraphics[width=0.84\columnwidth]{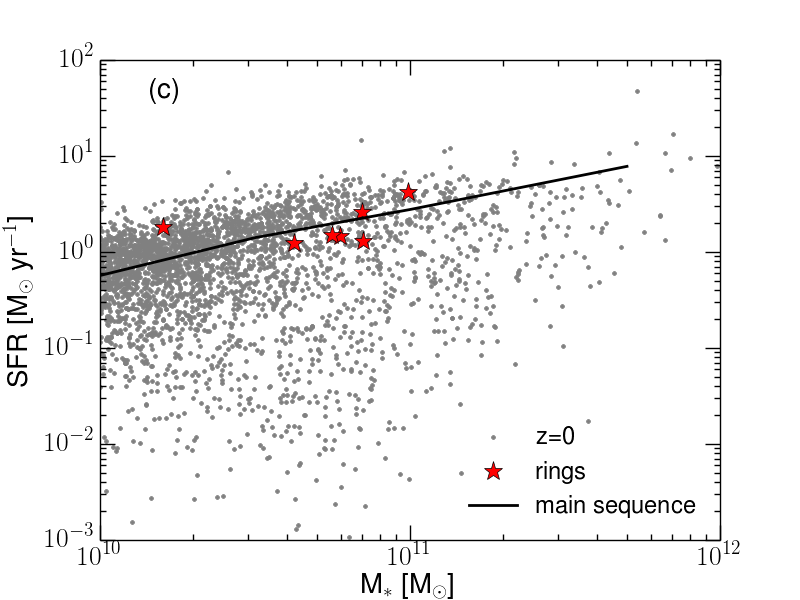}\includegraphics[width=0.84\columnwidth]{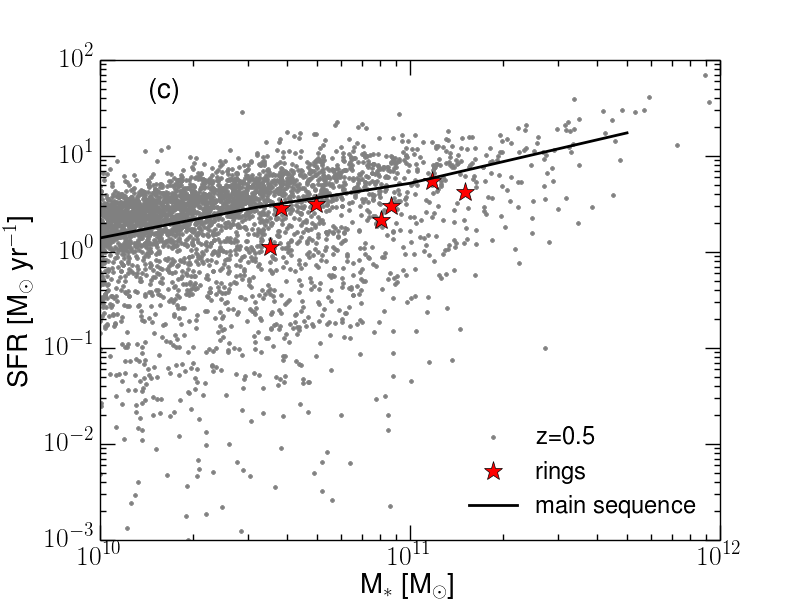}\\
\includegraphics[width=0.84\columnwidth]{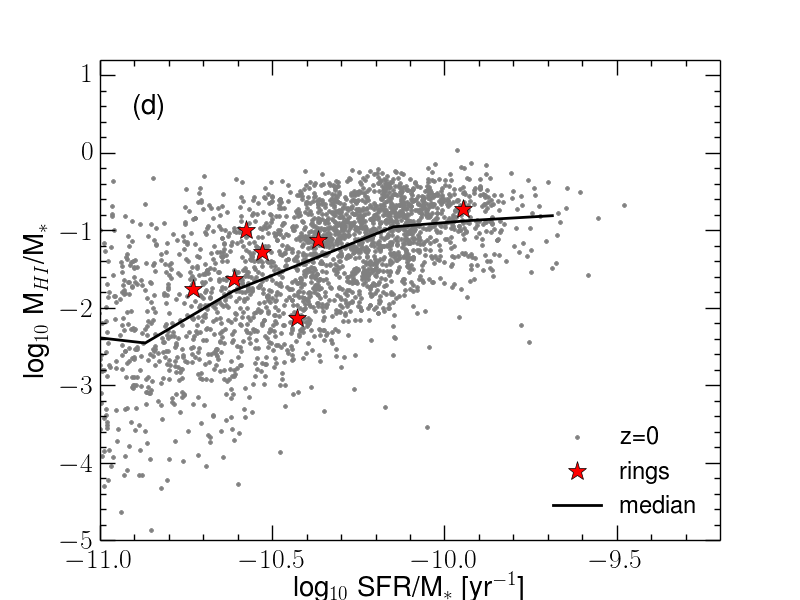}\includegraphics[width=0.84\columnwidth]{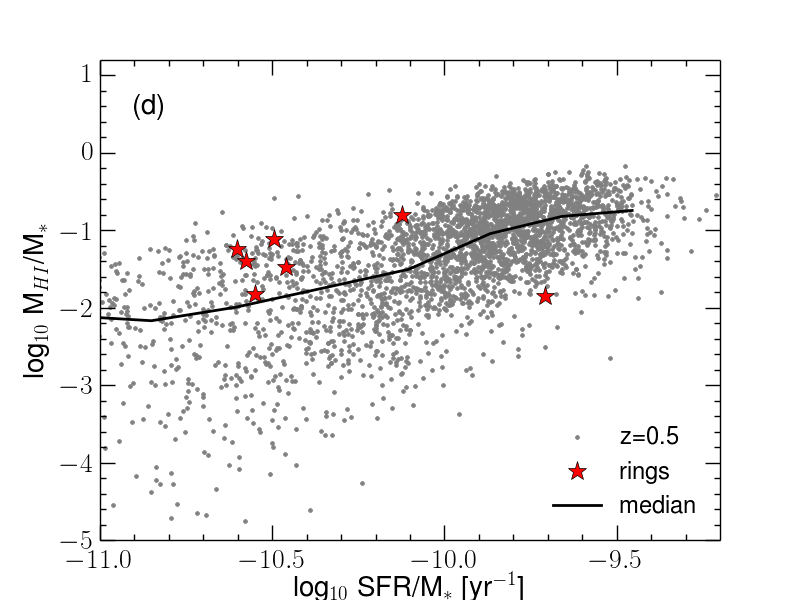}
\caption{Panels (a) show a scatter plot of EAGLE galaxies in the colour-stellar mass plane ($u-r$ vs. M$_{*}$) 
at $z\,=\,0$ (left) and $z\,=\,0.5$ (right). The red and blue lines mark the location of the red sequence and
blue cloud at each redshift \citep{2016Trayford}, respectively. 
Panels (b) present the HI gas fraction scaling relation (M$_{\x{HI}}$/M$_{*}$ vs. M$_{*}$) of
EAGLE galaxies (grey circles), and the black line marks the median gas fraction values at fixed stellar mass. 
Panels (c) show the star formation rate (SFR) versus the stellar mass for galaxies in EAGLE at
$z\,=\,0$ (left) and $z\,=\,0.5$ (right). The black line in this panel shows the star formation main 
sequence at each redshift \citep{2015-Furlong}. Panels (d) show the HI gas fraction (M$_{\x{HI}}$/M$_{*}$) versus the specific
star formation rate (SFR/M$_{*}$) for galaxies in EAGLE at $z\,=\,0$ (left) and $z\,=\,0.5$ (right).
The red stars in this figure represent the EAGLE ring galaxies at each redshift.}
\label{4e}
\end{figure*}

\subsection{The Properties of Eagle Ring Galaxies}
\noindent Figure \ref{4c} is a visualisation of fourteen randomly selected collisional ring galaxies in EAGLE in the redshift range between $z\,$=$\,0$ and $1.5$.
Each row corresponds to the face-on $u$-band mock image (left), the HI (middle) and H$_2$ (left) column density maps of
the same galaxy. The HI and the H$ 2$ column densities were calculated using the prescriptions in  \citet{2013-Rahmati} for the neutral hydrogen fraction in gas particles  and 
\citet{2011-Gnedin} for the HI to H$_2$ transition, as described in detail in \citet{2015-Lagos}. The colours in the column density images are coded according to the colour bar in the top row and
are in units of atoms per cm$^{2}$. The $u$-band images are $60\,$kpc on a side while the HI and H$_2$ column density images
are $100\,$kpc on a side. The atomic and molecular hydrogen column densities exhibit the same ring morphology seen in the $u$-band images,
and regions with very low column density ``holes'' are present in the HI and the H$_2$ maps around the nucleus and
the inner low surface brightness regions. This is in agreement with the current theoretical understanding of the formation history of
ring galaxies, in which the collision with a companion galaxy carries the gas of the target galaxy through  its disk  and forms 
the ring morphology. The majority of the observed collisional ring galaxies in the local Universe show similar morphologies, for instance the atomic gas in the Cartwheel and
AM0644-741 (Lindsay--Shapley Ring) is mostly concentrated in a ring morphology and a region of very low column density is clearly
visible in the nucleus; for reference see the HI maps of these two galaxies in \citet{Higdon-1996} and \citet{Higdon-2012}.\\

\begin{figure}
\centering
\includegraphics[width=1.0\columnwidth]{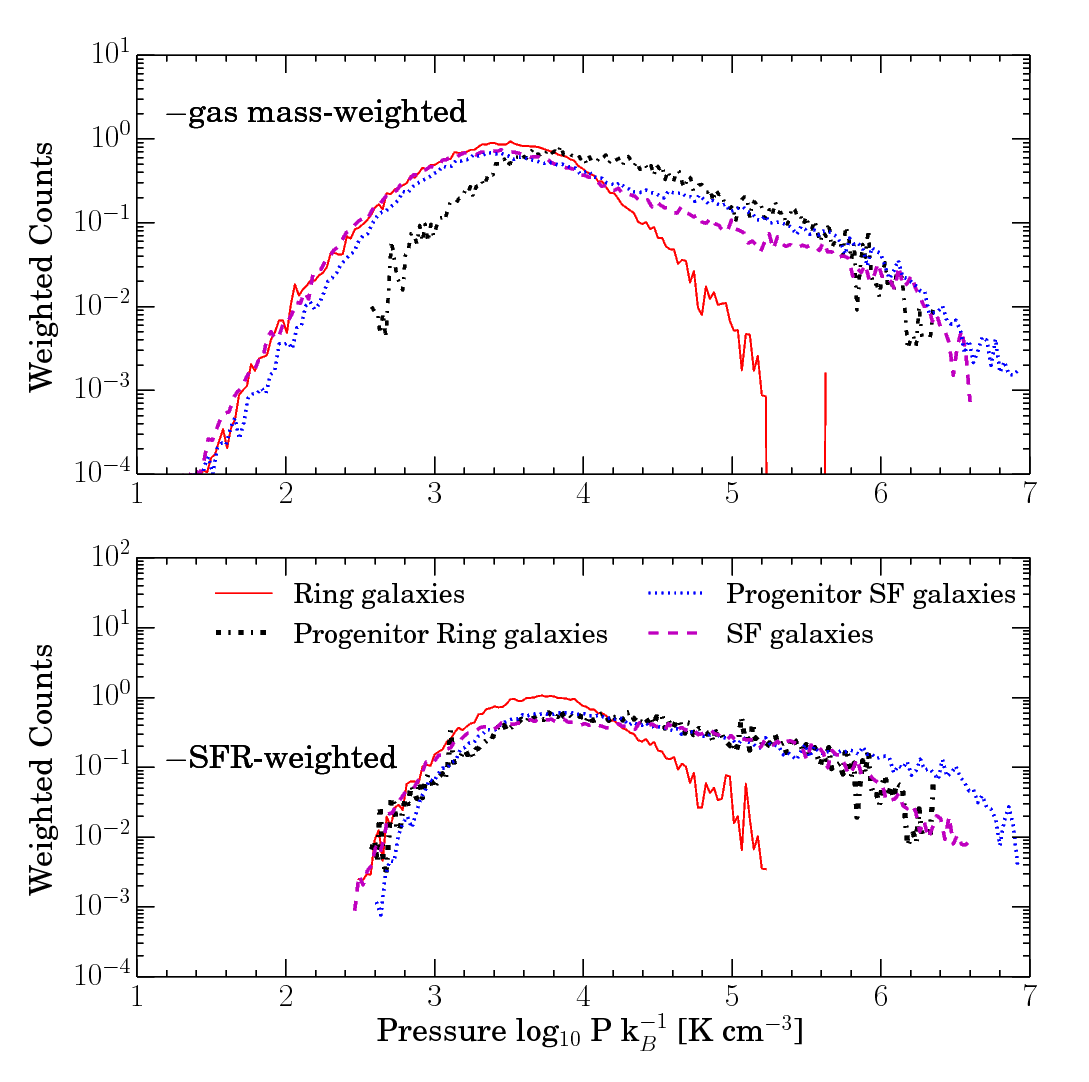}
\caption{The pressure of the gas particles in the EAGLE ring (red line) and control (magenta dashed line) galaxies at similar redshift expressed in 
terms of $P\, k^{-1}_{\x{B}}$, where k$_{\x{B}}$ is  Boltzmann's constant. The black dash-dotted and blue dotted lines show the pressure PDF of 
the ring and control galaxies two snapshots before ($z\,$=$\,0.7$), respectively. In the upper panel the pressure of the gas particles is normalised and
weighted by their neutral gas mass, whereas the lower panel presents the  pressure normalised and weighted by their SFR.}

\label{4d1}
\end{figure}

Figure \ref{fig_edge-on} presents the edge-on $gri$ colour composite images of a subsample of EAGLE ring galaxies identified 
between $z=0$ and $1$. The majority of the EAGLE rings have relatively thin stellar disks when viewed edge-on.
This is in agreement with non-cosmological isolated interaction simulations \citep{Gerber1996, Mapelli2012}, in which the rings in systems formed due
to collisions with  massive satellites (mass ratio M$_{\x{comp}}$/M$_{\x{ring}}\geq1.0$) contain larger fractions of their disk's material (gas and stars) and are
thicker in comparison with their counterparts which formed with less massive companions. Hence, EAGLE rings are expected to have relatively thin disks as the 
vast majority of these systems have less massive companions and stellar mass ratios less than one.
Figure \ref{fig_starview} shows the stellar volume density view in six ring galaxies identified in EAGLE between $z=0$ and $1$.
The volume density is calculated as the stellar mass in each cell divided by the cubic smoothing length, $\frac{4}{3}\pi h^3$,
with the image pixels smoothed over $1\,$kpc. In each view, a companion satellite galaxy appears within a close distance (roughly one ring diameter) 
from the central ring galaxy. We also note that the vast majority of EAGLE ring galaxies have companion satellite(s) that lie within similar distances to the examples in Figure
\ref{fig_starview}, on average within a distance  $\gtrsim100\,$kpc. This agrees with the current theoretical and observational
understanding of ring galaxies in the local Universe. For instance, in most of the observed ring galaxies the intruder lies within a physical
distance of less than $100\,$kpc \citep{Elagali2018,conn-2016,Fogarty-2011}.\\

Figure \ref{4e}(a) shows a scatter plot of EAGLE galaxies in the ($\,u\,-\,r\,$) colour-stellar mass plane
at $z\,=\,0$ (left) and $z\,=\,0.5$ (right). The colours here are intrinsic, i.e., rest-frame and dust-free colours; 
refer to \citet{2015-Trayford} for details on the dust modelling and the magnitude measurements. 
The red stars in this Figure correspond to ring galaxies in EAGLE at $z\,=\,0$ (left) and $z\,=\,0.5$ (right). 
The red and blue lines mark the location of the red sequence and blue cloud at each redshift, respectively, and are shown to highlight the colour bimodality of
the galaxies in EAGLE \citep{2016Trayford}. Most of the ring galaxies at $z\,=\,0$ are located in the
blue cloud  in the colour-stellar mass diagram which indicates that these systems are actively forming stars. This is comparable to the observations of
ring galaxies in which most of the studied collisional ring galaxies (within redshift range $z\,=0\,-0.1$) are star-forming blue
galaxies  \citep{Wong2006, Romano2008,Fogarty-2011,Parker-2015,conn-2016,Elagali2018}. However, at $z\,=\,0.5$ ring galaxies are located in the green 
valley with fewer candidates in the blue cloud. This suggests that the interactions driving the ring morphology are also driving the colour transformation  and gas 
exhaustion which may lead to quenching in these galaxies.\\

Figure \ref{4e}(b) presents the standard HI gas fraction scaling relation (M$_{\x{HI}}$/M$_{*}$vs.\,M$_{*}$)
of the EAGLE galaxies at $z\,=\,0$ (left) and $z\,=\,0.5$ (right). 
At both redshifts, the majority of the ring galaxies have higher gas fraction
in comparison with the median at fixed stellar mass. This highlights the high HI gas fraction in collisional ring galaxies found in the EAGLE simulations, 
which agrees well with the observations, in which collisional ring galaxies are known to have high HI gas fractions relative to other galaxies at
fixed stellar mass \citep{Elagali2018}. Figure \ref{4e}(c) shows the SFR vs. the stellar mass of EAGLE galaxies at $z\,=\,0$ (left) and $z\,=\,0.5$ (right). 
The black line shows the star formation main sequence at each redshift \citep{2015-Furlong}. At redshift $z\,=\,0$, the majority of ring galaxies are  active 
star-forming galaxies and lie on the main sequence with few occupying the green valley. 
On the other hand, at redshift $z\,=\,0.5$ ring galaxies have considerably lower star formation rates at fixed stellar mass than the main 
sequence. This is consistent with the top right panel of the same Figure, where ring galaxies appear to be in the process of quenching and occupy the green
valley in the colour-stellar mass plane. It is important to note that massive galaxies in EAGLE (M$_{*}\gtrsim10^{10}\,$M$_{\odot}$) are known to have slightly
less HI gas than in observations \citep{2017Crain}, which is expected to have a direct impact on the star formation rates resulting from the interaction.
This can be part of the reason why ring galaxies in EAGLE have less star formation rate in comparison with the observations, see for example the early IRAS study of ring 
galaxies in \citet{Appleton-Struck-Marcell1987}. We study the ISM of ring galaxies in the next subsection to explore any other physical reasons that may lead to the low star 
formation in ring galaxies in EAGLE. Figure \ref{4e}(d) shows the HI gas fraction (M$_{\x{HI}}$/M$_{*}$) versus the specific star formation rate 
for galaxies in the EAGLE simulation at $z\,=\,0$ (left) and $z\,=\,0.5$ (right). The black   
line marks the median values at each sSFR bin. At redshift $z\,=\,0.5$, the majority of the
ring galaxies have a higher gas fraction than the median value at their sSFR bin. This is different for ring galaxies at redshift
$z\,=\,0$ as only two rings have higher gas fraction at fixed sSFR.\\

\begin{figure}
\centering
\includegraphics[width=1.0\columnwidth]{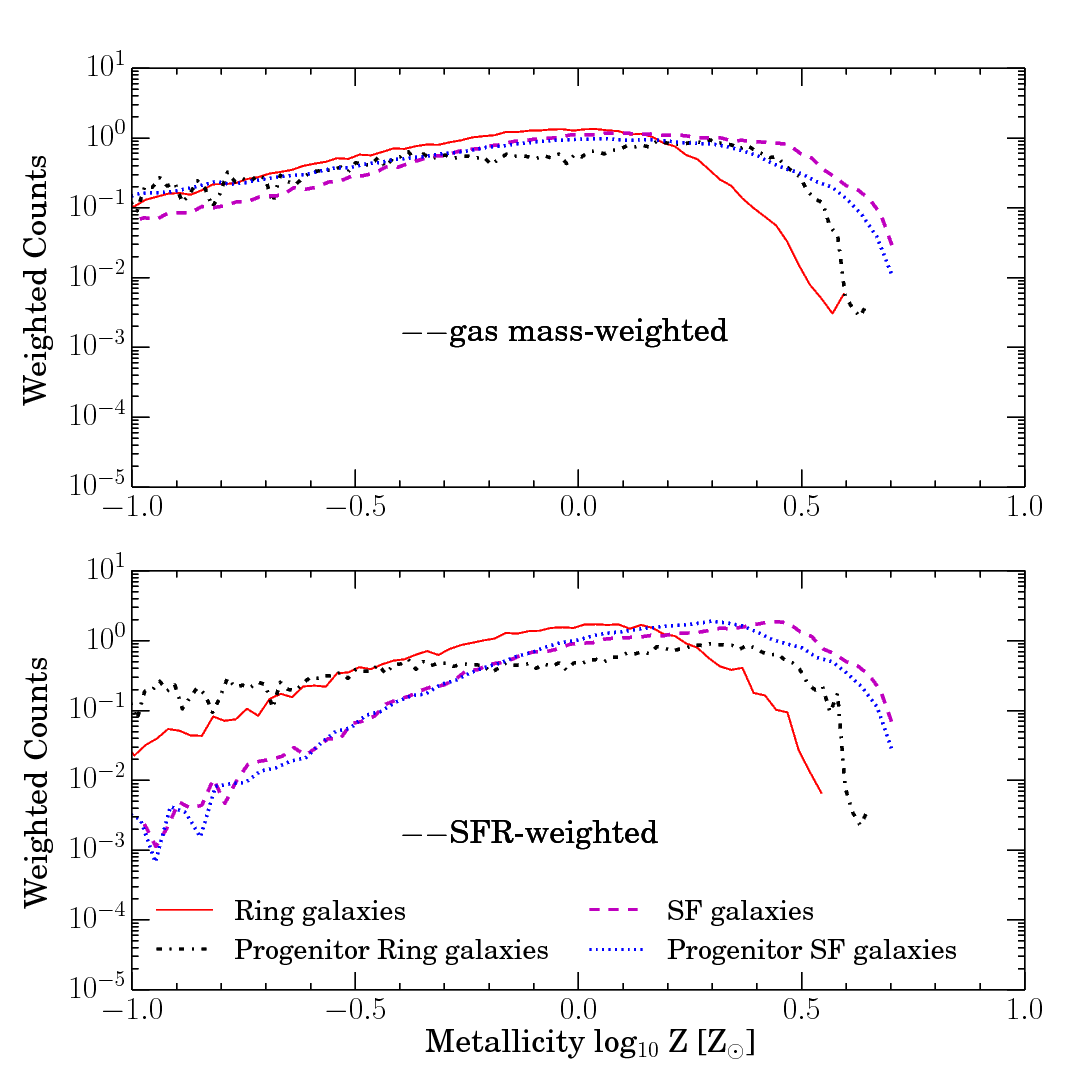}
\caption{The metallicity of the gas particles in the EAGLE ring (red line) and control (magenta dashed line) galaxies expressed in units of solar metallicities. 
 The black dash-dotted and blue dotted lines show the metallicity PDF of the  progenitor ring and control galaxy samples at $z\,$=$\,0.7$, respectively. 
In the upper panel the metallicity of the gas particles is normalised and weighted by their neutral gas mass, whereas the lower panel presents the  metallicity normalised and weighted by their SFR.}
\label{4d2}
\end{figure}

It is important to mention that we explored these standard scaling relations up to $z=1.5$, however we only show two redshifts for brevity.
Ring galaxies at higher redshifts have the same trend as rings at $z=0.5$; the majority of these systems have high gas fractions yet 
lie below the star formation main sequence and occupy the green valley in the colour-stellar mass plane. 
Combining the results of the HI and the SFR of ring galaxies versus the general population of galaxies in EAGLE, we see that ring galaxies
are characterised by an inefficient conversion of the HI gas into stars. It is important to caution the readers to the 
low number statistics in this analysis; even though it is evident that rings at higher redshifts ($z\gtrsim0.5$) exhibit inefficiency 
in star formation these systems maybe unrepresentative of the whole population of ring galaxies at higher redshifts.
We discuss the physical drivers behind this inefficiency in Section 4.4.\\

\subsection{The ISM of Eagle Ring Galaxies}
EAGLE adopts the star formation law of \citet{2008Schaye}, which expresses the \citet{Kennicutt-1998} observational relation as a pressure law, with the SFR of an 
individual gas particle scaling as P$^{n-1/2}$, with n=$\,1.4$. The gas particles, however, are only assigned a SFR if they reach a 
certain density level, which depends on the metallicity of the gas \citep{2004Schaye}. This means that lower pressure directly translates into a lower SFR. 
To better understand the ISM of ring galaxies and the reasons behind the high amount of HI gas and the inefficient star formation in these system,
we explore the properties of the gas particles in seven ring galaxies at redshift $z\,$=$\,0.5$ and compare  them with seven other EAGLE star-forming galaxies with
similar gas and stellar masses. We refer to the latter as the control galaxy sample. The average difference in the HI gas mass between the ring and control samples
is smaller than $0.05$ dex. \\

\begin{figure}
\centering
\includegraphics[width=1\columnwidth]{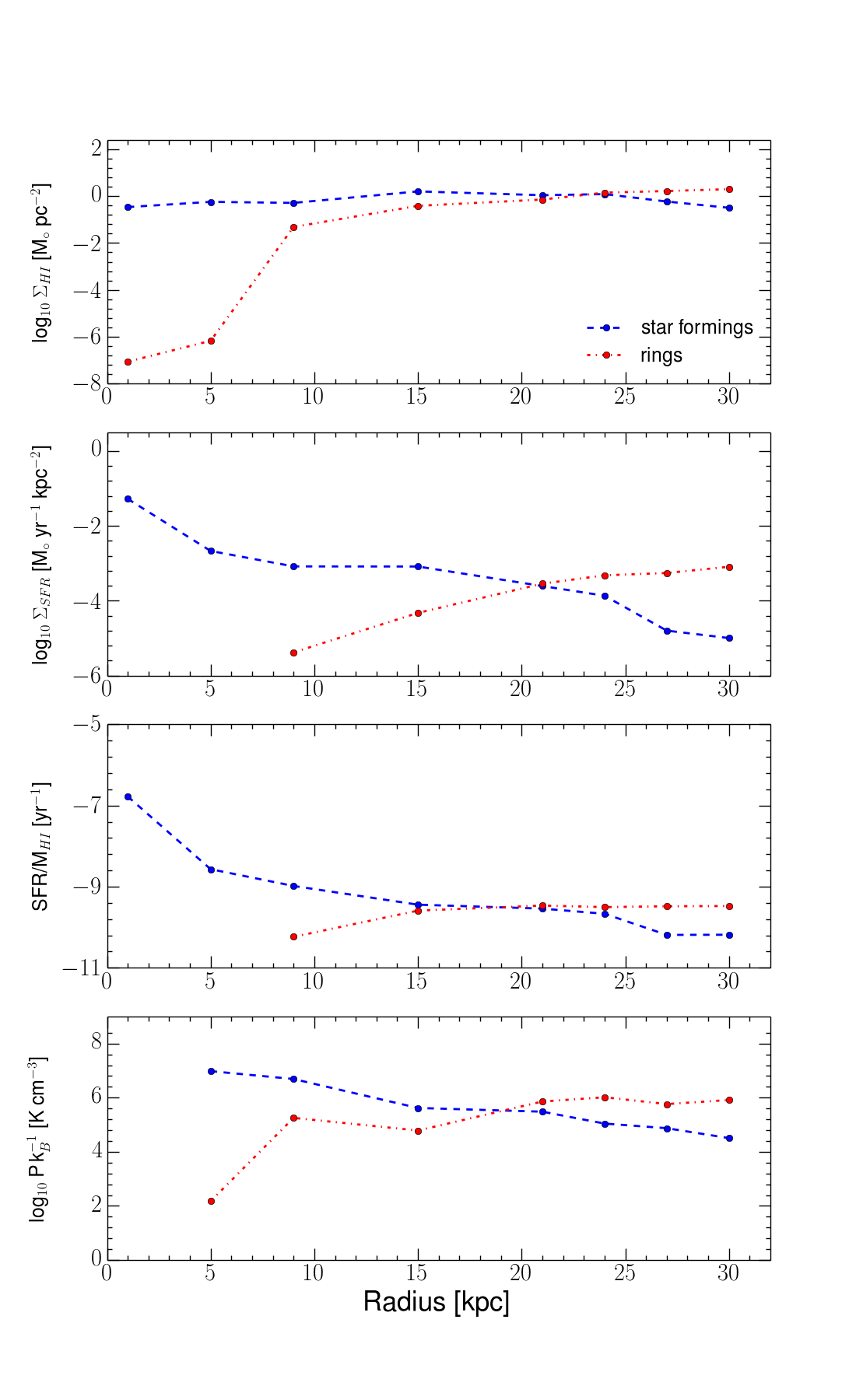}
\caption{The median radial profile of the HI gas surface density $\Sigma_{\x{HI}}$ (top panel), the star formation rate surface density 
$\Sigma_{\x{SFR}}$ (second panel), the specific star formation rate SFR/M$_{\x{HI}}$ (third panel), and the hydrostatic pressure of the gas (bottom panel) in  the control 
seven star-forming galaxies (blue) and EAGLE ring galaxies (red) at $z\,=\,0.5$.}
\label{4d3}
\end{figure}

Figure \ref{4d1} shows the pressure of the gas particles in ring (red line) and control (magenta dashed line) galaxies at the same redshift expressed in 
terms of $P\, k^{-1}_{\x{B}}$, where k$_{\x{B}}$ is  Boltzmann's constant. The black dash-dotted and blue dotted lines show the pressure profile of 
the ring and control galaxies two snapshots before the ring feature is identified ($z\,$=$\,0.7$), respectively. We refer to those 
as the progenitors of ring and control galaxies. Here, we adopt a $50\,$kpc spherical aperture centred on
the subhalo's centre of mass and consider all the material within this aperture to belong to the subhalo (the galaxy).
The upper panel of Figure \ref{4d1} shows the pressure of the gas particles weighted by their neutral gas masses, whereas 
the lower panel presents the same property weighted by SFR.\\

There is a clear difference in the gas pressure between the ring and the control (star-forming) galaxy samples.
The gas in the progenitor of both the ring and control galaxies, as well as and control galaxies at $z\,$=$\,0.5$
have higher pressure regardless of the weighting. This is one reason for the inefficient star formation processes taking place 
in ring galaxies and can partly explain the excess amount of HI gas in these system. 
This is because the gas-phase pressure affects the rate at which the HI gas is converted to molecular hydrogen, which physically is expected to 
regulate the conversion into stars \citep{1989-Elmegreen,Elmegreen-1994,Elmegreen-1993,Blitz-2006, Leroy-2008}, and in some models is the only
parameter controlling  whether molecular or atomic gas dominates the ISM in galaxies \citep{Lagos-2011}. 
This means that high column density, low-pressure HI gas has relatively few H$_2$ molecules and therefore a low SFR in comparison with high column density, high-pressure
HI gas. We note that the gas-phase pressure of the particles evolves with redshift, see Figure 12 in \citet{2015-Lagos}, which is the reason for the 
shift in the pressure profile of the two progenitor samples at $z\,$=$\,0.7$ towards higher values overall compared to
the ring and control galaxy samples at $z\,$=$\,0.5$, especially in the upper panel.\\

Figure \ref{4d2} shows the metallicity of the gas particles in the previously discussed seven rings (red line) and control (magenta dashed line) galaxy samples
expressed in units of solar metallicities. The black dash-dotted and blue dotted lines present the metallicity profile of the  progenitors of the ring 
and control galaxies at $z\,$=$\,0.7$, respectively. The upper panel of Figure \ref{4d2} shows the metallicity of the gas
particles weighted by their neutral gas masses, whereas the lower panel presents the metallicity weighted by their SFR. 
It is evident from  Figure \ref{4d2} that the gas particles in ring galaxies have lower metallicities compared to the
control galaxy sample. It is important to note that the difference in the metallicity profile between the ring galaxies and 
their progenitors is likely due to the interaction with the dwarf satellite in which the ring galaxies may gain low metallicity
gas from the dwarf as a result of the interaction. \\

\begin{figure*}
\centering
\includegraphics[width=0.7\columnwidth]{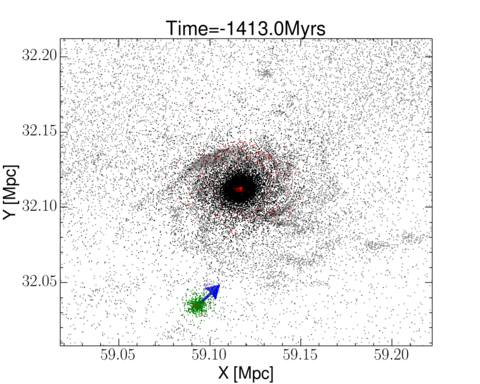}\includegraphics[width=0.7\columnwidth]{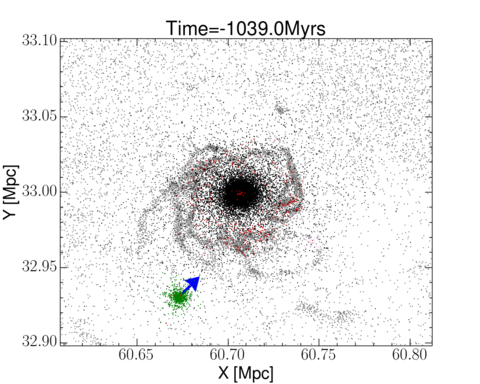}\includegraphics[width=0.7\columnwidth]{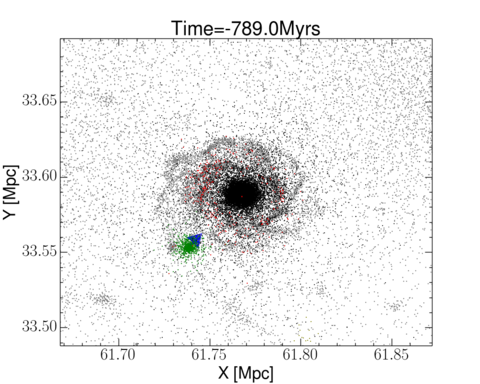}\\
\includegraphics[width=0.7\columnwidth]{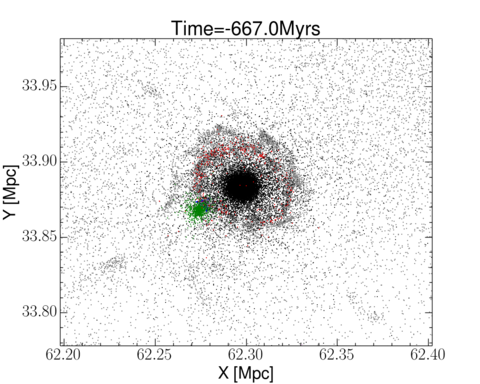}\includegraphics[width=0.7\columnwidth]{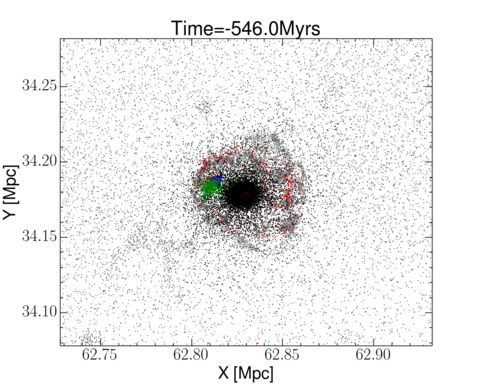}\includegraphics[width=0.7\columnwidth]{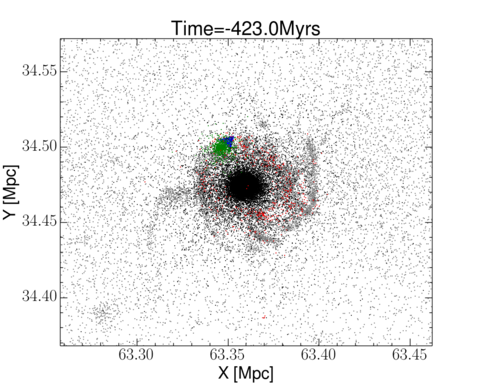}\\
\includegraphics[width=0.7\columnwidth]{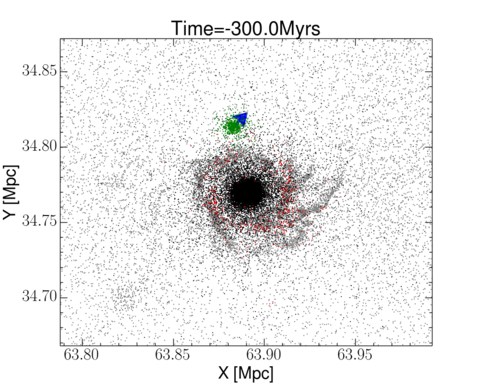}\includegraphics[width=0.7\columnwidth]{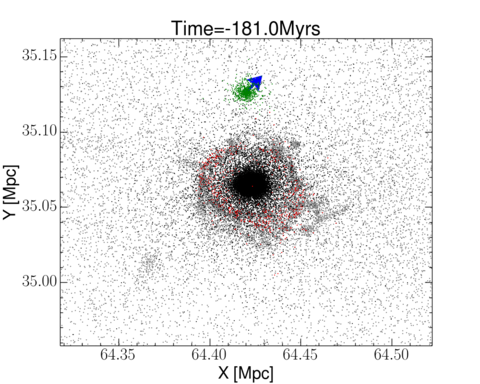}\includegraphics[width=0.7\columnwidth]{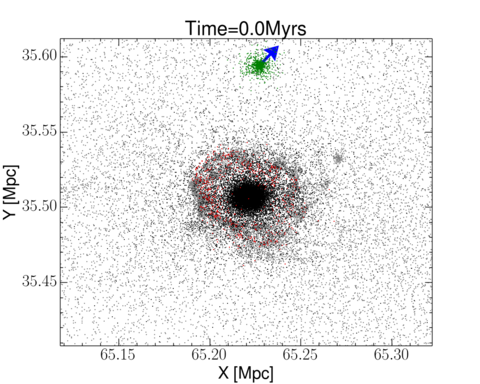}\\
\caption{The morphological evolution  of a gas-rich spiral in EAGLE colliding once with a dwarf galaxy. In all the maps the system is viewed down the simulation
z-axis. The grey dots show the gas particles, while the red and black dots show newly formed stars (age < $30\,$Myr) and those with age > $30\,$Myr, respectively.
The green dots show the stellar particles of the dwarf companion. This system is identified as a ring galaxy at $z\,=\,0$. The time of the snipshot in Megayears with respect
to the ring formation time ($z\,$=$\,0$) is denoted at the top of each panel. The blue arrows show the magnitude and the direction of the velocity vector of the dwarf galaxy.
We print half the number of particles to avoid overcrowding.}
\label{figpp}
\end{figure*}

An intriguing question to address is whether the overall low gas-phase pressure in ring galaxies is due to the different gas distribution in these systems, in which most
of the gas piles up in regions of low pressure, or whether the gas-phase pressure in these galaxies is low irrespective of radii.
In the latter case, the pressure at a given radius would be lower than that of the control sample. We explore these two scenarios
in Figure \ref{4d3}. This figure presents the median radial profile of the HI gas surface density $\Sigma_{\x{HI}}$ (top panel), the star formation rate 
surface density $\Sigma_{\x{SFR}}$ (second panel), the specific star formation rate SFR/M$_{\x{HI}}$ (third panel), and the hydrostatic pressure of the 
gas (bottom panel) in  the control seven star-forming galaxies (blue) and the ring galaxies (red) at $z\,$=$\,0.5$.
It is important to note that these ring galaxies have different physical disk radii, hence their ring features are not necessarily at the same radius.
The two different samples have similar neutral hydrogen gas surface densities at radii $r > 5\,$kpc but differ dramatically at radii $r\lesssim5\,$kpc.
This is connected to the suppression of the star formation and pressure in the inner regions of ring galaxies when compared to the 
control sample. However, this suppression takes place out to larger radii than the depression in the HI surface density of ring galaxies.
These trends show that both scenarios described above are taking place in our ring galaxies in EAGLE.
At radii  $r\,>\,20\,$kpc, ring galaxies have higher hydrostatic pressure in comparison with their counterparts. The drop in the gas-phase pressure seen
in the inner regions of ring galaxies is caused by the drop-through collision, which leads to a ``dilution'' in 
the ISM gas surface density and as a consequence the star formation surface density drops significantly in the inner radii of these galaxies.
This is evidenced by the progenitors of ring galaxies having the same pressure distributions as the progenitors of the control sample.\\

The relatively low pressure and metallicity in ring galaxies offers a physical interpretation for
the inefficient HI to SFR conversion, mostly in the inner regions of these systems, which can lead to the high HI gas fraction
found in observations of ring galaxies \citep{Elagali2018}. Further, this can also explain the low total molecular hydrogen masses
in some of the observed ring galaxies in the local Universe. For instance, the Lindsay-Shapley ring galaxy has as much total H$_{2}$ mass as a 
typical dwarf galaxy \citep{2005Leroy,2010Higdon}, which is surprising especially because this galaxy has an atomic gas mass 
of M$_{\x{HI}}\,$=$3\times10^{10}\,$M$_{\odot}$ \citep{Higdon-2012}.
Recent observational studies of ring galaxies show that some of these galaxies are H$_{2}$ deficient even in the outer ring where the atomic
hydrogen surface density is highest. The most plausible scenario offered in these studies is that the molecular hydrogen in the ring is 
destroyed by ultraviolet photons from OB stars born in the ring's confined environment \citep{2015Higdon, 2017Wong}.
This is different than the result presented in Figure \ref{4d3}, where the star formation rate surface density in ring galaxies is higher in the outer 
radii ($r > 20\,$kpc) in comparison with the star-forming galaxy sample implying that the abundance of cold gas in the ring should be relatively normal.
To examine the  photodissociation hypothesis suggested in \citet{2015Higdon, 2017Wong}, 
we need to follow the formation and the evolution of the molecular hydrogen. This requires a detailed description of the ISM and the photoionisation effects of local sources
on the H$_{2}$ gas, which is not currently feasible in simulations with a boxsize as large as EAGLE and only prescribed in zoom-in simulations,
see for example \citet{Hopkins2014,Hopkins2017}. However, future developments in large hydrodynamical simulations, such as EAGLE, will explicitly track the formation of the molecular hydrogen at much higher 
resolution. Such projects will be ideal to address this issue.\\

\begin{figure*}
\centering
\includegraphics[width=0.7\columnwidth]{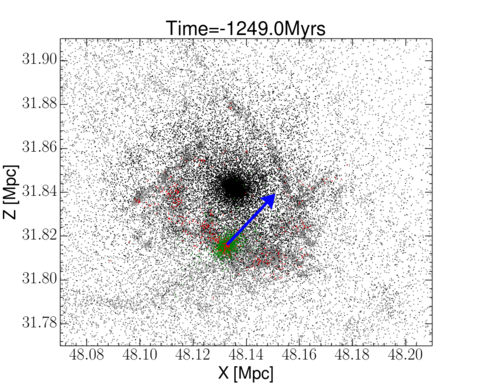}\includegraphics[width=0.7\columnwidth]{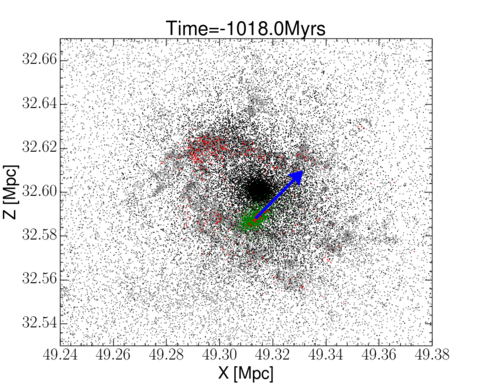}\includegraphics[width=0.7\columnwidth]{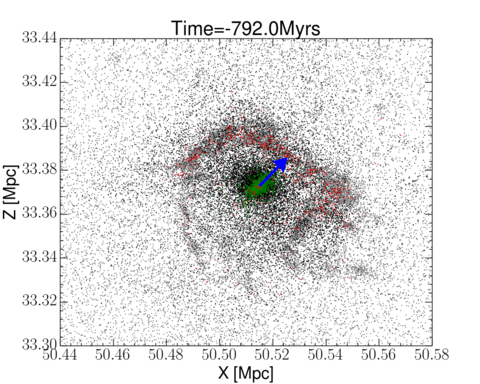}\\
\includegraphics[width=0.7\columnwidth]{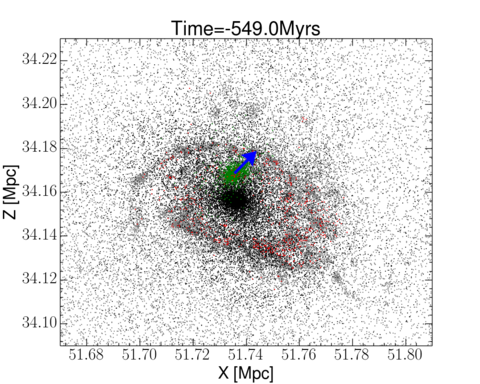}\includegraphics[width=0.7\columnwidth]{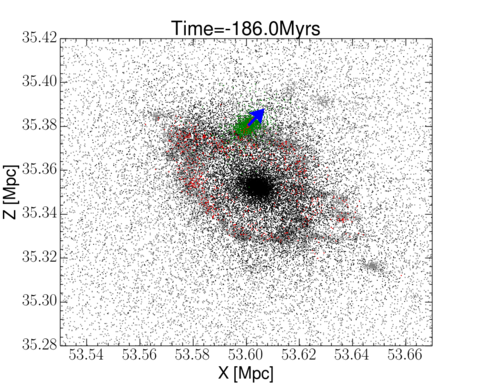}\includegraphics[width=0.7\columnwidth]{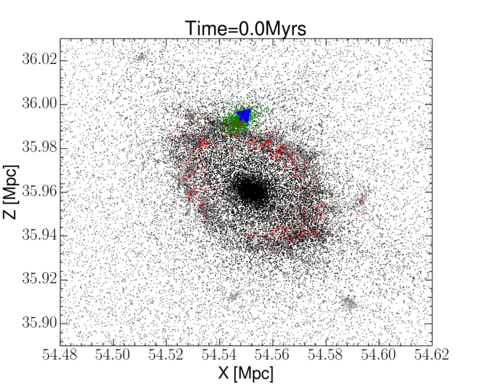}\\
\includegraphics[width=0.7\columnwidth]{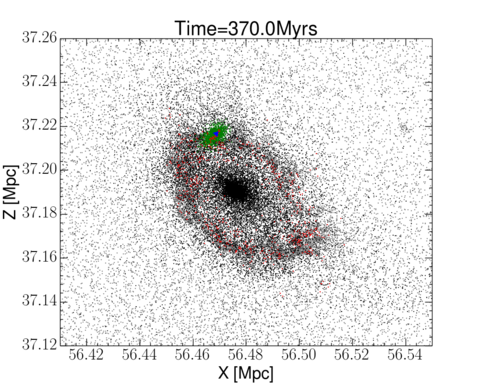}\includegraphics[width=0.7\columnwidth]{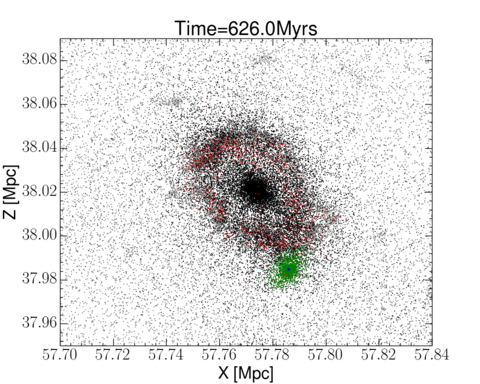}\includegraphics[width=0.7\columnwidth]{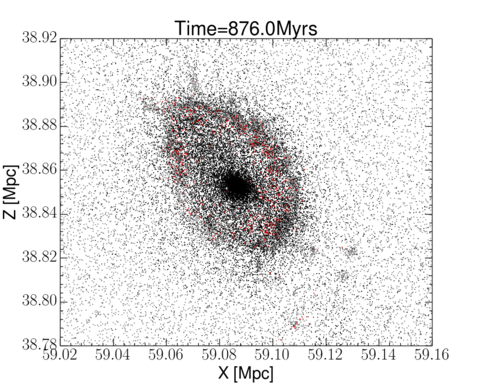}\\
\caption{The morphological evolution  of a gas-rich spiral in EAGLE colliding twice with a dwarf galaxy. In all the maps the system is viewed down the simulation
y-axis. The gas and stars are colour-coded similar to Figure \ref{figpp}. This system is identified as a ring galaxy at $z\,=\,0.5$. The time of the snipshot in Megayears with respect
to the ring formation time ($z\,$=$\,0.5$) is denoted at the top of each panel. The blue arrows show the magnitude and the direction of the velocity vector of the dwarf galaxy.
We print half the number of particles to avoid overcrowding.}
\label{figpp2}
\end{figure*}

\begin{figure}
\centering
\includegraphics[width=0.48\columnwidth]{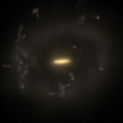}\includegraphics[width=0.48\columnwidth]{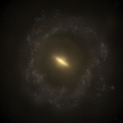}\\
\includegraphics[width=0.48\columnwidth]{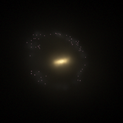}\includegraphics[width=0.48\columnwidth]{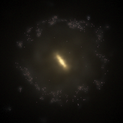}\\
\includegraphics[width=0.48\columnwidth]{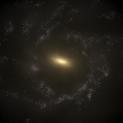}\includegraphics[width=0.48\columnwidth]{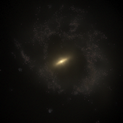} 
\caption{Three-colour $gri$ mock images of a subsample of the long-lived EAGLE ring galaxies identified in  the redshift range between $z=0$ and $1.0$. 
These images are $60\,$kpc on a side and are available at the EAGLE database webpage \citep{2016McAlpine}.}
\label{bars}
\end{figure}

\begin{figure}
\centering
\includegraphics[width=1.0\columnwidth]{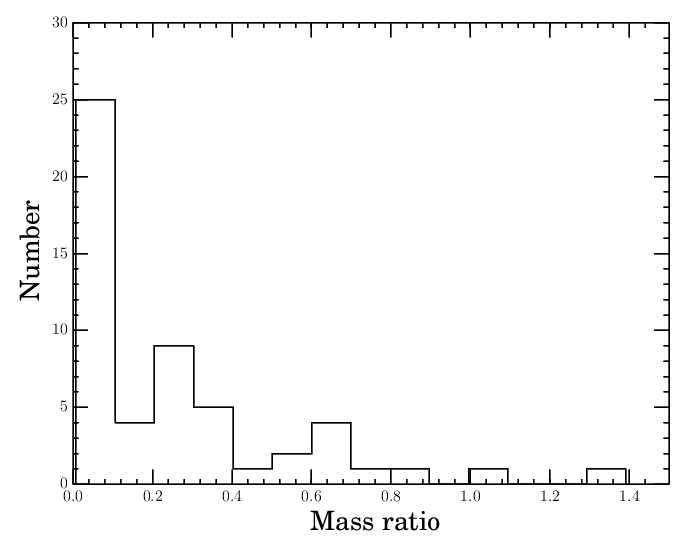}
\caption{The stellar mass ratio between the companion and the ring galaxies for our sample of rings in EAGLE that formed due to collisions. Here, we show all rings at  $z\,\leq\,1.5$.}
\label{5b}
\end{figure}

\section{Formation Mechanisms of Eagle Ring Galaxies}
To understand the formation mechanisms of ring galaxies in EAGLE, we trace the evolution of the haloes hosting 
ring galaxies using the finer time resolution snipshots \citep{Schaye-2015}. The time resolution of these snipshots 
span between $0.05-0.3\,$Gyr, which is smaller than the typical merger timescale \citep{2014Ji,2008Jiang}, hence
very suitable to track collisions and drop-through interactions. For each ring galaxy, we examine the history of their
host halo by tracing its evolution  for one Gigayear before and one Gigayear after the ring morphology is identified.\\

\begin{table}
\centering
\caption{The formation mechansims of ring galaxies in the EAGLE simulations.}
\label{my-label2}
\begin{tabular}{lll}\hline \hline
 Mechanism & Number & Averaged lifespan (Myr) \\ \hline 
One collision encounter &  $37$&  $700$\\
Multiple collision encounters  &$9$  &$1100$  \\
Barred galaxies (P-type) & $9$ &$2400$   \\ \hline
\end{tabular}
\end{table}

\begin{figure}
\centering
\includegraphics[width=1.0\columnwidth]{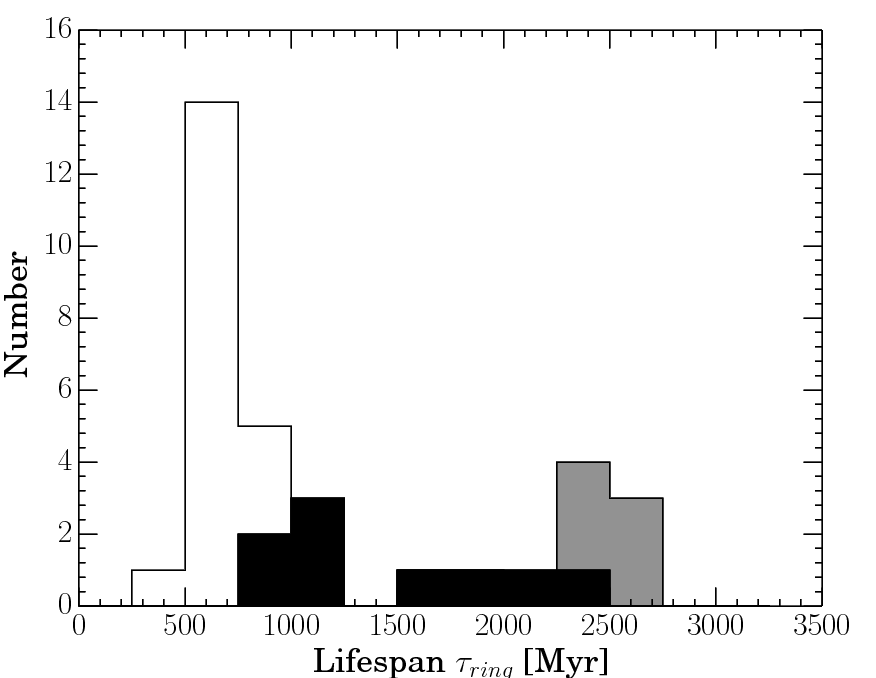}
\caption{The lifetime of the ring of star formation morphology in our EAGLE ring galaxies at $z\leq1.5$.
The unfilled histogram shows the lifespan of the ring galaxies formed with one companion dropping once through its gas disk, while the 
black filled histogram show those which have companion(s) plunging through more than once. 
The grey histogram represents the long-lived ring galaxies, those galaxies have very strong bars.}
\label{5c}
\end{figure}

We visually inspect the history of the all the ring galaxies identified in EAGLE to determine whether the ring morphology
forms in response to collisions or not. This process is very simple, as in all the cases the effect of the drop-through is evident if present. 
An example of this is shown in Figures \ref{figpp}-\ref{figpp2} (discussed in detail below).
The majority of ring galaxies identified in EAGLE ($46$ out of $55$ galaxies) have an interaction origin, i.e., formed when a companion
galaxy(ies) drop-through a "target" galaxy (sometimes more than one drop-through). The remainder  galaxies ($17\,$per cent)
have very long-lived ring morphologies ($> 2\,$Gyr). These long-lived systems are most likely barred galaxies (P-type rings). 
For instance, one of these long-lived ring galaxy hosts, identified at $z\,=\,0$, is listed as a strongly barred galaxy in A17.
Table 2 lists the different identified formation mechanisms for ring galaxies in EAGLE and the average lifespan of the ring morphology
in each scenario. Figure \ref{figpp} presents the interaction history of a gas-rich, spiral colliding with a dwarf companion galaxy at redshift $z\,$=$\,0$. 
The ring morphology in this scenario is induced due to a single off-centre collisional encounter with a dwarf companion. As listed in Table 2,
the majority of ring galaxies in EAGLE are formed in this scenario. Figure \ref{figpp2} presents the second formation scenario for
ring galaxies in EAGLE at redshift $z\,$=$\,0.5$. In this case, a gas-rich, bulgeless spiral galaxy collides with a dwarf galaxy multiple
times, i.e., more than one drop through collision. This second scenario is expected since the intruder and the target galaxy are more bound after the first encounter as a result of the dynamical
friction. In EAGLE, this scenario often prompt a full merger after a few hundred megayears from the second encounter.
Multiple encounters can prolong the lifespan of the ring and results in different kinematics for the gas and stars of target galaxy (see Figure \ref{velocityexpansion}),
the lifespan of the ring morphology for these scenarios is shown in Figure \ref{5c}
(discussed in detail below). Figure \ref{bars} shows a subsample of the P-type (long-lived) ring galaxies identified in EAGLE at $z\leq1.5$.
The ring morphology in this class is mainly due to the strong bars present in these galaxies.
These systems have ring features that last for $> 2\,$Gyr and host strong bars that are evident in the $gri$ mock images of this figure.\\

Figure \ref{5b} shows the stellar mass ratio between the companion and the ring galaxy (M$_{\x{comp}}$/M$_{\x{ring}}$) in the
$84\,$ per cent of our sample of ring galaxies, in which we find a clear connection between the ring morphology and galaxy interactions.
The median stellar mass ratio in this sample equals $0.14$, which is in agreement with the observational and theoretical studies of these systems. 
The stellar mass ratios in most of the observationally studied ring galaxy pairs lie between $0.1$ to $1$  
\citep[see for example][]{Higdon1997,Higdon1995, Wong2006,Romano2008, Parker-2015, 2017Wong}. Only a handful of ring galaxy pairs
have ratios outside this range. For instance, three known ring galaxies have mass ratios smaller than $0.1$, namely 
IZw45 (ratio$\sim 0.04$), NGC 2793 (ratio$\sim0.05$) and NGC 922 (ratio$\sim0.06$), while three others have mass ratios larger
than one in which the companion is twice the size of the ring galaxy, namely, the Arp 141, Arp 147 and Arp 148 \citep{Romano2008}.
In EAGLE, we identify two ring galaxies that have more massive companions. The interaction in these two cases is not a ``bullseye'' collision but rather 
a collision with large impact parameter, i.e. similar to a flyby interaction.\\

Figure \ref{5c} also shows the lifetime of the star-forming ring morphology found in our  ring galaxy sample.
The lifetime of the ring is determined from  when the collision occurs between the target and companion galaxies in the case of collision induced rings, 
i.e., from the time when the companion is closest to the target until the time when the ring structure starts to collapse and hosts less than $50$ per 
cent of the new stars ($< 30\,$Myr) formed in the whole galaxy. The unfilled histogram shows the ring structure lifespan of the ring galaxies 
formed with one companion dropping once through its gas disk, while the black filled is for those which have multiple companions ($\,>\,1$) plunging
through more than once (predominantly identified at higher redshifts). For the latter, the lifetime is measured from the first collision.
The grey histogram presents the long-lived ring galaxies found in EAGLE. The median lifespan of the ring morphology in systems formed with one companion dropping once through its disk is 
$700\,$Myr. Multiple interactions, i.e. more than one companion passing through the disk of the main galaxy or the same companion plunging more than once, 
significantly prolong the lifespan of the ring feature in the main galaxy.\\

\begin{figure}
\centering
\includegraphics[width=1\columnwidth]{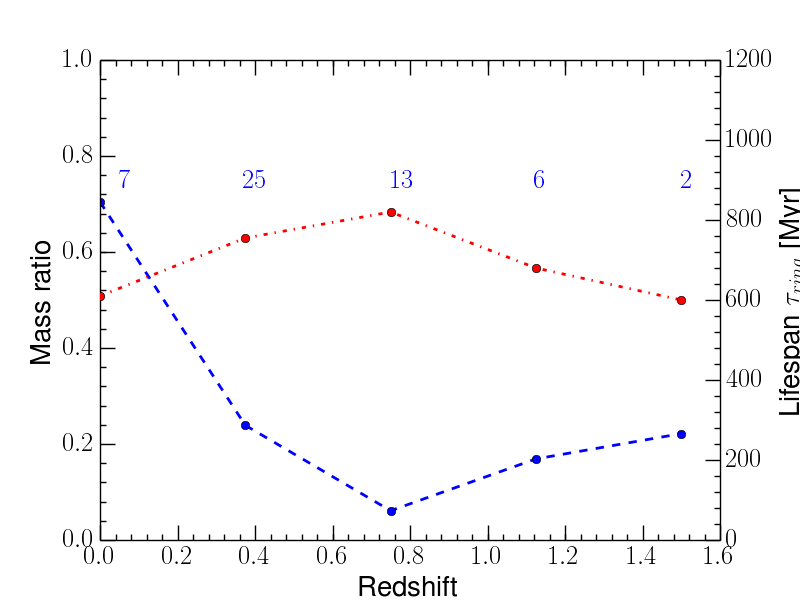}
\caption{The evolution of the median stellar mass ratio of the interacting galaxy pairs (blue) and the median lifetime of the ring morphology (red).
Here, we only show rings that have an interaction origin and exclude the long-lived barred ring systems. Numbers represent the total number
of ring galaxies within each redshift bin.}
\label{5d}
\end{figure}

This value is in broad agreement with the ages of the ring morphology in non-cosmological isolated interaction simulations, which in most  
cases range between $500\,$Myr and $\,1000\,$Myr \citep{Hernquist-1993, Mihos-1994,2001Horellou,MapelliMoore-2008}. Figure \ref{5d} shows the evolution of the median stellar 
mass ratio of the interacting galaxy pairs (blue) and the median lifetime of the ring morphology (red) with redshift. 
Here, we only show rings that have an interaction origin and exclude the long-lived barred ring systems. 
The numbers in this Figure represent the total number of ring galaxies within each bin. The stellar mass ratio of the interacting pairs are the highest at $z=0$, and decrease rapidly with 
redshift. On the other hand, the lifetime of the ring morphology increases monotonically with redshift until it reaches the maximum at  $z\,$=$\,0.75$, 
with a median age of $\sim800\,$Myr, and then decreases at higher redshifts.
The evolution of the mass ratio is expected as galaxies tend to be more gas rich and smaller at higher redshifts, and consequently the life time of the 
ring morphology will also evolve. However, the steep drop in the mass ratio can also result from small number statistics.
The next  generation of hydrodynamical simulations  will offer a chance to revisit this with better statistics, as it will simulate galaxies 
in larger volumes, allowing the exploration of redshift trends more robustly.\\

Figure \ref{velocityexpansion} presents the radial velocity of the gas particles with respect to the plane of the disk for two  
ring galaxies in EAGLE at redshift $z\,$=$\,0.5$. The plane of the disk is determined as that perpendicular to the total stellar spin vector of the galaxy. We only 
consider the component of the gas particles in the radial direction of the plane of the disk. Thus, the radial velocity is a measure of the expansion velocity of the ring.
Figure \ref{velocityexpansion}a shows the radial velocity of a ring that formed due to one drop-through interaction with a dwarf satellite and is measured at time 
$t\sim200\,$Myr after the encounter. The radial velocity of the ring at $r\,\gtrsim\,17\,$kpc is positive and indicates expansion,
while in the inner regions the velocity is negative which means that the gas is falling back to the centre.  For instance, the  outer rings have a maximum 
radial velocity of $v_{\x{rad}}\,$=$120\,$km s$^{-1}$ at a radius of $r\,$=$\,25\,$kpc, whereas at a radius of $r\,$=$\,7\,$kpc the radial velocity equals
$v_{\x{rad}}\sim-160\,$km s$^{-1}$. This result is in agreement with the predictions of the analytic caustic theory \citep{Appleton1996,Struck-Marcell-Lotan1990,Struck2010,Mapelli2012} and 
the observations of ring galaxies in the local Universe \citep{conn-2016, Fogarty-2011, Higdon-1996}.
Figure \ref{velocityexpansion}b shows the radial velocity profile of a long lived ring galaxy formed due to two encounters  with a dwarf companion 
and is measured at time $t\sim140\,$Myr after the second encounter. The radial velocity profile in this case is different 
than in Figure \ref{velocityexpansion}a in that the inward radial velocity (infall) in the inner regions of the ring are much smaller in this case and only particles at radii $r\gtrsim 5\,$kpc
have negative velocities. Assuming that the second collision is also impulsive, we expect the velocity of the gas particles to undergo a dramatic change, in which the velocity of the 
second impulse will be added to the gas particles' current  velocity. This means that the particles moving inward (negative infall velocity) will gain 
momentum which will reduce the net inward negative velocity or even cancel it, while the gas particles moving outward (expanding)
will gain more speed and will increase their expansion velocity (refer to \citet{Appleton1996,Struck-Marcell-Lotan1990} for more detailed analysis on multiple encounters).

\section{Discussion and Conclusions}
In this paper, we use the EAGLE hydrodynamical simulations  to study the formation and characteristics of ring galaxies.
EAGLE's  volume, ($100$)$^{3}$ cMpc$^{3}$,  allows us to identify a relatively large sample of ring galaxies
and quantify their formation mechanisms and evolution with redshift. We characterise our ring galaxy sample, putting special emphasis on their 
star formation rates (SFRs), colours, metallicities, and atomic (HI) hydrogen gas and how these properties scale with each other. We also characterise
their environments. Our main findings are summarised as follows:

\begin{figure}
\centering
\includegraphics[width=1.0\columnwidth]{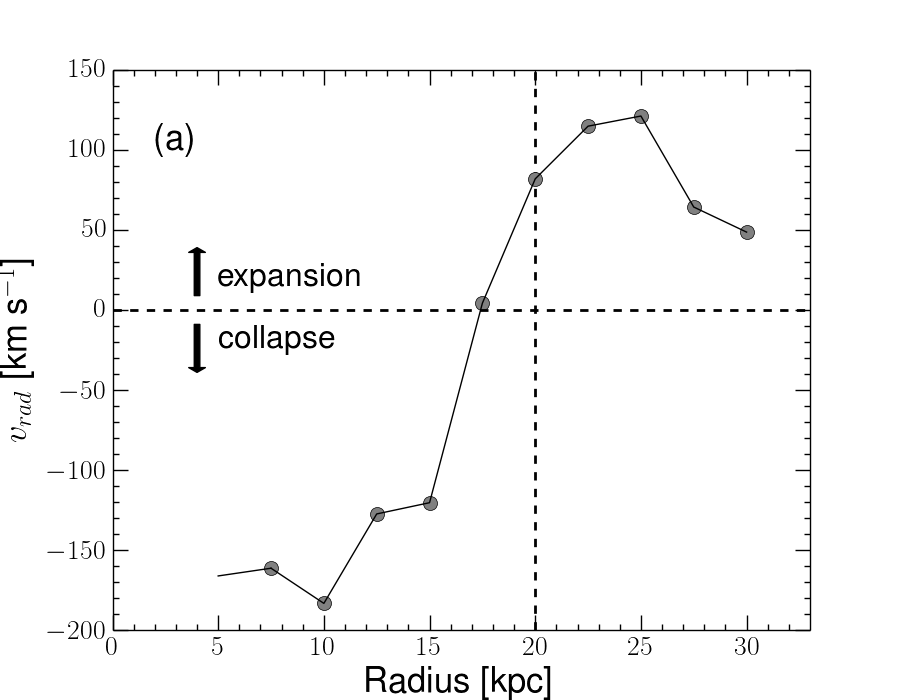}\\
\includegraphics[width=1.0\columnwidth]{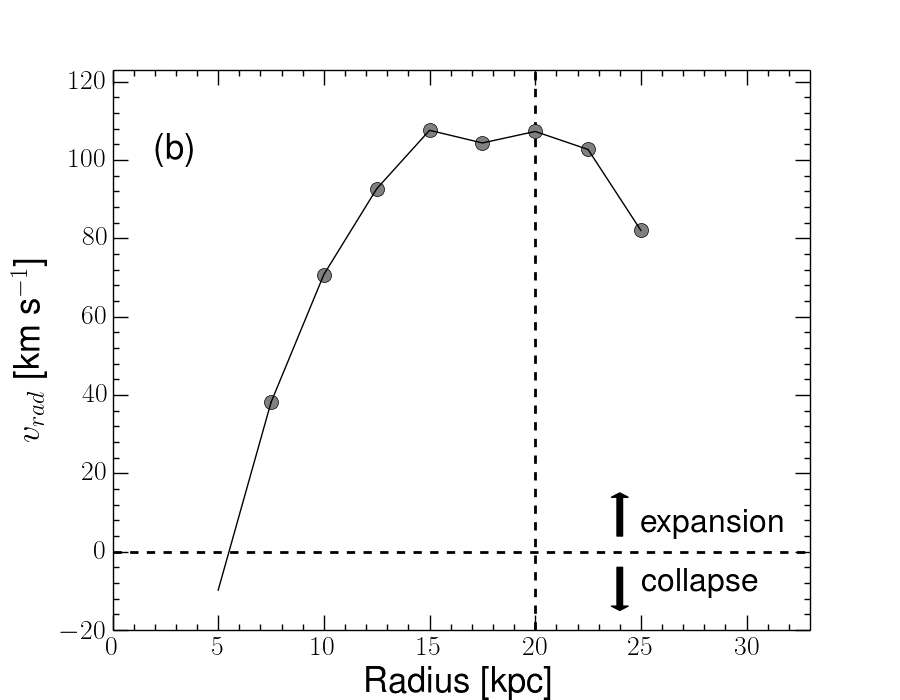}\\ 
\caption{The radial velocity of the gas particles in the plane of the disk for two different EAGLE ring galaxies 
at redshift $z\,$=$\,0.5$. (a) a ring galaxy formed due to one drop-through interaction with a dwarf satellite. (b) a ring 
galaxy formed due to multiple drop-through interactions with a dwarf companion (two encounters). The radial velocity is measured after $t\sim200\,$Myr
from the first encounter in (a) and after $t\sim140\,$Myr from the second encounter in (b). The dashed vertical line denotes the approximate location of the ring.}
\label{velocityexpansion}
\end{figure}

\begin{itemize}

 \item The number density evolution of ring galaxies in the EAGLE simulation is in broad agreement with the observations
 of \citet{Lavery-2004,Elmegreen-2006}. This means that the  numerical treatment of the ISM, star formation, and 
 feedback model in the EAGLE simulation are adequate enough to reproduce a realistic ring-morphology population. 
 This is an unprecedented success for hydrodynamical simulations  especially because the calibration of the subgrid 
 physics in EAGLE does not include galaxy morphology \citep{Crain-2015}. This is particularly important as other simulations
 produce an overabundance of ring galaxies \citep[e.g., the Illustris simulations][]{Snyder-2015}.\\

\item Ring galaxies live in massive groups (M$_{\x{halo}}\sim10^{13}\,$M$_{\odot}$) that are preferentially 
 more concentrated than groups without ring galaxies at  fixed halo mass. This is in agreement with the observations,
 in which ring galaxies are located within galaxy groups and have at least one companion galaxy
 \citep{Higdon1997,Higdon1995,Romano2008,conn-2016,2017Wong,Elagali2018}.\\
 
\item In EAGLE, ring galaxies are moderately star-forming galaxies that typically reside in the green valley, but have high HI gas fractions.
We find that this is due to ring galaxies having an ISM with much lower gas phase-pressure and metallicity than star-forming galaxies with 
the same stellar and gas masses. By studying the progenitors of ring galaxies, we find that the drop-through collision with the companion(s) 
is responsible for diluting the gas metallicity and pressure. The latter happens as the gas flows efficiently towards the outskirts of galaxies 
to form the ring structure, where the pressure is lower.\\

\item The vast majority of ring galaxies identified in EAGLE ($83\,$per cent) have an interaction origin, i.e., are formed when 
a companion galaxy(ies) drop-through a "target" galaxy (sometimes more than one drop-through). The lifespan of the 
ring morphology in systems formed with one companion dropping once through its disk is $\sim700\,$Myr, which is  is in broad
agreement with the ages of the ring morphology in non-cosmological isolated interaction simulations \citep{Hernquist-1993, Mihos-1994,2001Horellou,MapelliMoore-2008}.
Some ring galaxies form through the effect of more than one companion passing through its disk or the same companion passing more than once .
In these cases, we find that the lifespan of the ring feature increases significantly, by a factor of $\sim2$.
We also study the  kinematics of EAGLE collisional ring galaxies and find similar results to those predicted by the analytic 
caustic theory \citep[see e.g.,][]{Appleton1996,Struck-Marcell-Lotan1990}. The remainder  galaxies ($17\,$per cent) have very long-lived ring morphology ($> 2\,$Gyr), and correspond to barred galaxies (P-type rings).

\end{itemize}

One aspect of these systems remains unclear mainly due to the limitation of the current hydrodynamical simulations.
Some recent observations suggest that ring galaxies are H$_2$ deficient even in the ring, where the HI gas surface density is highest, and hypothesise
that in the ring's confined high density ISM, molecular hydrogen is formed due to the high HI surface density but equally destroyed by the 
continuous UV-field from supernovae  and OB stars \citep{2015Higdon, 2017Wong}. To explore the photodissociation effect in the ring's molecular hydrogen, we need 
to consistently follow the formation and destruction  of the molecular hydrogen, which requires a detailed description of the cold phase of the ISM.
This is currently not available for large hydrodynamical simulations such as EAGLE, but in smaller box zoom-in simulations is possible
and has been implemented to some extent \citep{Hopkins2014,Hopkins2017}.\\

Another  limitation of our current analysis is the small number statistics. Ring galaxies are rare systems, as a result we typically find less than ten rings in the
entire simulated volume  at a given redshift. Thus, any redshift evolution found here is tentative and requires larger volumes, simulated at the same resolution to EAGLE,
to confirm them. Using the next generation of hydrodynamical simulations, we will be able to probe larger volumes 
at the same resolution as the current EAGLE simulations, which will improve our statistics and allow us to explore redshift trends more accurately. \\

With the upcoming large sky surveys and telescope missions, such as the James Webb Space Telescope \citep[JWST;][]{2006Gardner,2018Kalirai}, 
the MeerKAT Karoo Array Telescope  HI Surveys \citep{Holwerda}, the Australian Square Kilometre Array Pathfinder HI surveys 
\citep{2007Johnston, 2008Johnston}, the Multi-Unit Spectroscopic Explorer \citep[MUSE;][]{2010Bacon}, the K-band Multi-Object Spectrograph 
\citep[KMOS;][]{2015Wisnioski} surveys, and Atacama Large Millimeter/submillimeter Array (ALMA) future surveys,
we expect to be able to study ring galaxies at higher redshifts and larger volumes due to their deep and  large sky coverage. Through these surveys, we will also be able to test 
the simulation predictions and probe both the gas-phase pressure and the metallicity in these galaxies. This will further advance our understanding of the ISM as 
well as the star formation law in extreme collision cases.\\

\section*{Acknowledgements}
We thank the referee, Curt Struck, for his comments which significantly improved the presentation of this manuscript.
We also thank Joop Schaye for useful comments on the manuscript, and David Algorry for providing the list of barred galaxies identified in EAGLE and published in \citet{Algorry-2017}.
AE  wishes to Acknowledge the funds he received from the International Centre of Radio Astronomy (ICRAR).  
CL is funded by an Australian Research Council Discovery Early Career Researcher Award (DE150100618) and by the
Australian Research Council Centre of Excellence for All Sky Astrophysics in 3 Dimensions (ASTRO 3D), through project 
number CE170100013. MS is supported  by  VENI  grant 639.041.749. We acknowledge the Virgo Consortium for making their simulation data available. The EAGLE simulations
were performed using the DiRAC-2 facility at Durham, managed by the ICC, and the PRACE facility Curie based in France at TGCC,
CEA, Bruyeresle-Chatel. This work used the DiRAC Data Centric system at Durham University, operated by the Institute for Computational Cosmology on behalf 
of the STFC DiRAC HPC Facility ({\tt www.dirac.ac.uk}). This equipment was funded by BIS National E-infrastructure capital grant ST/K00042X/1, STFC capital grant ST/H008519/1,
and STFC DiRAC Operations grant ST/K003267/1 and Durham University. DiRAC is part of the National E-Infrastructure.

\bibliographystyle{mnras.bst}
\bibliography{myref}

\appendix
\section{List of EAGLE ring galaxies}
Table \ref{my-labele} lists all the  ring galaxies identified in the EAGLE simulations up until redshift $z$=$\,2.23$.

\begin{table}
\centering
\caption{Ring galaxies identified in the EAGLE simulation.} 
\label{my-labele}
\begin{tabular}{|lllll|}\hline 
 GalaxyID & Snapshot & z& Group N.& Subgroup N. \\ \hline
1419938&&&164&1\\ 
13176886&&&63&2\\ 
14042156&&&130&1\\ 
15358400&28&0&258&0\\ 
15511663&&&281&0\\ 
15978820&&&342&0\\
18010353&&&841&0\\ \hline
16736005&&&501&0\\ 
16750450&&&467&0\\ 
14042157&&&465&0\\ 
11525816&27&0.10&15&5\\ 
13839082&&&181&2\\ 
8903544&&&1174&0\\ 
15476546&&&281&0\\ 
20755891&&&120&0\\ \hline
8930054&&&1232&0\\ 
9078223&&&994&0\\ 
13176888&26&0.18&62&1\\ 
14681036&&&206&0\\
16832146&&&32&1\\ 
18010355&&&841&0\\  \hline
18010356&&&842&0\\ 
16347102&&&532&0\\ 
13825639&25&0.27&164&2\\ 
20629170&&&128&0\\  \hline
13660663&&&76&2\\
15978824&&&540&0\\
16229689&&&668&0\\
16482632&24&0.36&432&0\\
16921472&&&630&0\\
17451600&&&519&0\\
17961354&&&557&0\\ \hline
14418325&&&473&0\\ 
15187269&&&845&0\\ 
16482633&23&0.50&478&0\\ 
15511668&&&342&0\\ 
18026177&&&888&0\\ 
15978825&&&569&0\\ 
20259289&&&107&0\\ \hline
18078238&&&1133&0\\ 
15528603&22&0.61&477&0\\ 
16521451&&&500&0\\ \hline
15528604&&&476&0\\ 
15289527&21&0.73&403&0\\ 
8931586&&&1333&0\\ \hline
15528605&20&0.86&531&0\\ 
16701972&19&1.0&913&0\\ 
14371833&19&1.0&130&0\\ 
9135999&19&1.0&1805&0\\
16701973&18&1.26&965&0\\
17643587&18&1.26&905&0\\
17643588&17&1.48&828&0\\
19899640&17&1.48&352&0\\ 
20640573&16&1.73&233&0\\
20727481&14&2.23&780&0\\ \hline
\end{tabular}
\end{table}

\label{lastpage}

\bsp

\end{document}